\documentclass[revtex4-2,twocolumn,showpacs]{openjournal}

\def \be{\begin{equation}}
\def \ee{\end{equation}}
\def \bdm{\begin{eqnarray}}
\def \edm{\end{eqnarray}}
\usepackage{float}
\usepackage{amsmath}
\usepackage{graphicx}
\def \d{\partial}

\begin{document}

\title{Subspace Approximations to the Focused Transport Equation of Energetic Particles \\I. The Standard Form}

\author{B. Klippenstein and A. Shalchi}
\affiliation{Department of Physics and Astronomy, University of Manitoba, Winnipeg, Manitoba R3T 2N2, Canada}
\email{andreasm4@yahoo.com}
\begin{abstract}
The Fokker-Planck equation describing the transport of energetic particles interacting with turbulence is difficult to solve analytically. Numerical solutions are of course possible
but they are not always useful for applications. In the past a subspace approximation was proposed which allows to compute important quantities such as the characteristic
function as well as certain expectation values. This previous approach was applied to solve the one-dimensional Fokker-Planck equation which contains only a pitch-angle
scattering term. In the current paper we extend this approach in order to solve the Fokker-Planck equation with a focusing term. We employ two- and three-dimensional subspace
approximations to achieve a pure analytical description of particle transport. Additionally, we show that with higher dimensions, the subspace method can be used as a
hybrid analytical-numerical method which produces an accurate approximation. Although the latter approach does not lead to analytical results, it is much faster compared to
pure numerical solutions of the considered transport equation.
\end{abstract}
%
%

\maketitle

\section{Introduction}\label{sec:intro}

Energetic particles such as cosmic rays interact with large scale magnetic fields and turbulence. The latter interaction is complicated and leads to a diffusive motion
of particles described by diffusion coefficients (see, e.g., Shalchi (2009a) and Shalchi (2020) for reviews). However, non-diffusive transport has been discussed in
the literature (see, e.g., Zimbardo (2005), Zimbardo et al. (2006), Perri \& Zimbardo (2007), Shalchi \& Kourakis (2007), Perri \& Zimbardo (2009a), Perri \& Zimbardo (2009b),
Tautz \& Shalchi (2010), Zimbardo et al. (2012), and Perri et al. (2015)).

For the mean magnetic field it is often assumed that it is constant and points into the $z$-direction. In this case the mean field has the form $\vec{B}_0 = B_0 \hat{z}$
where $B_0$ is a constant. However, real magnetic fields are not constant. Parker (1958), for instance, suggested that the solar magnetic field can be described by an
\textit{Archimedean spiral}. To include such a complicated mean magnetic field into analytical theories for energetic particle transport is very difficult and only numerical
test-particle simulations can be performed for this case (see Tautz et al. (2011)).

The standard approach to incorporate a varying mean magnetic field is via the focusing effect originally introduced into the field of particle transport by Roelof (1969).
This effect leads to an additional term in the transport equation used to describe the particle motion. As describe in Schlickeiser (2002) and Zank (2014), there  is a certain
hierarchy of transport equations. At the most fundamental level one can use a relativistic version of the Vlasov equation (see Vlasov (1938)). The next level provides an equation
for the ensemble averaged distribution function in phase-space. In the context of cosmic rays this is often called the \textit{Fokker-Planck equation} (see Schlickeiser (2002)).
The third level is obtained by pitch-angle averaging the Fokker-Planck equation leading to a diffusive transport equation (see, e.g., Parker (1965)).

The general Fokker-Planck equation of cosmic ray transport is lengthy and contains several terms such as stochastic acceleration and perpendicular diffusion
(see, e.g., Skilling (1975), Schlickeiser (2002), and Zank (2014)). In the current paper we consider the following transport equation
\be
\frac{\partial f}{\partial t} + v \mu \frac{\partial f}{\partial z}
= \frac{\partial}{\partial \mu} \left[ D_{\mu\mu} \big( \mu \big) \frac{\partial f}{\partial \mu} \right] - \frac{v}{2 L} \big( 1 - \mu^2 \big) \frac{\partial f}{\partial \mu}
\label{FPeq}
\ee
where we have used the Fokker-Planck coefficient of pitch-angle scattering $D_{\mu\mu}$. The latter scattering parameter is a function of the pitch-angle cosine $\mu$.
The second term on the right-hand-side is the focusing term. The parameter $L$ is assumed to be constant and is called the \textit{focusing length}. It corresponds
to a characteristic length scale over which one observes a variation of the mean magnetic field. The focusing length is usually introduced by considering a mean
magnetic field where the $z$-component is given by
\be
B_{0z} \left( z \right) = B_0 e^{-z/L}
\label{B0zfocused}
\ee
where we have used the constant $B_0$. It follows from the solenoidal constraint $\vec{\nabla} \cdot \vec{B}_0 = 0$ and the assumption of an axi-symmetric magnetic field that
\be
B_{0x} \left( x, z \right) = \frac{x}{2 L} B_{0z} \left( z \right) \;\;\textnormal{and}\;\; B_{0y} \left( y, z \right) = \frac{y}{2 L} B_{0z} \left( z \right).
\label{B0xyfocused}
\ee
Eq. (\ref{FPeq}) is also known as the \textit{focused transport equation}. It was explored in several papers over the past few decades (see, e..g, Earl  (1976), Kunstmann (1979), Ruffolo (1995),
Shalchi (2009b), Shalchi (2011b), Shalchi \& Danos (2013), Danos et al. (2013), Wang \& Qin (2019), and Wang \& Qin (2020)). The equation investigated here neglects the effects of
a finite non-zero solar wind speed. The focused transport equation for relativistic flows was discussed by Webb (1985) as well as Webb (1987). A version of the focused transport equation for
relativistic background plasma flows was obtained Kirk et al. (1988) who discussed the effects of backward and forward moving Alfv\'en waves on the energization of the particles due to second
order Fermi acceleration.

It is the purpose of the current paper to solve Eq. (\ref{FPeq}) by using the $N$-dimensional subspace method described in Lasuik \& Shalchi (2019) and Shalchi (2024).
In Section 2 we discuss the subspace method by including the focusing term. Thereafter we approximately solve the Fokker-Planck equation analytically for $N=2$ in Section 3.
After finding the solution, several quantities such as expectation values are computed in Section 4. In Section 5 we discuss the accuracy and speed of the $N$-dimensional
subspace approximation method. Limitations of the proposed method are also discussed. In Section 7 we summarize and conclude.

\section{Fourier Transform and Subspace Approximation}

In the following we discuss the $N$-dimensional subspace method for the focused transport equation. In order to solve Eq. (\ref{FPeq}) we first rewrite it by using a Fourier
transform of the form
\be
f \big( z, \mu, t \big) = \int_{-\infty}^{+\infty} d k_{\parallel} \;  F_{k_\parallel} \big(\mu, t \big) e^{i k_{\parallel} z}.
\label{FourierTransform}
\ee
The corresponding inverse transform is given by
\be
 F_{k_\parallel} \big(\mu, t \big) = \frac{1}{2 \pi} \int_{-\infty}^{+\infty} d z \; f \big( z, \mu, t \big) e^{- i k_{\parallel} z}.
\label{invFourier}
\ee
Using this in Eq. (\ref{FPeq}) gives us
\bdm
& & \frac{\partial F_{k_\parallel}}{\partial t} + i v \mu k_{\parallel} F_{k_\parallel} \nonumber\\
& = & \frac{\partial}{\partial \mu} \left[ D_{\mu\mu} \frac{\partial F_{k_\parallel}}{\partial \mu} \right] - \frac{v}{2 L} \big( 1 - \mu^2 \big) \frac{\partial F_{k_\parallel}}{\partial \mu}.
\label{FourierFPeq}
\edm
The latter equation can be attempted to be solved via the expansion
\be
F_{k_\parallel}(\mu,t) = \sum_{n=0}^{\infty} C_n(t) P_n (\mu)
\label{expandF}
\ee
where we have used the \textit{Legendre Polynomials} $P_n (\mu)$. Furthermore, the yet unknown coefficients $C_n(t)$ are functions of time.
The idea of expanding the solution of the pitch-angle scattering equation into eigenfunctions, in our case the Legendre polynomials, was used
before in the literature (see, e.g. Earl (1974)). In such previous work the intention was to include the effects of cosmic ray inertia, the role of coherent
particle pulse propagation, and the effects of dispersive waves on the particle transport.

Using the expansion given by Eq. (\ref{expandF}) in Eq. (\ref{FourierFPeq}) yields
\bdm
& & \sum_{n} \dot{C}_n P_n + i v \mu k_{\parallel} \sum_{n} C_n P_n \nonumber\\
& = & \sum_{n} C_n \frac{\partial}{\partial \mu} \left[ D_{\mu\mu} \frac{\partial P_n}{\partial\mu} \right]\nonumber\\
& - & \frac{v}{2 L} \big( 1 - \mu^2 \big) \sum_{n=1}^{\infty} C_n \frac{\partial P_n}{\partial \mu}
\label{FokkerPlanckwithexpand}
\edm
where $\dot{C}_n$ denotes the time-derivative of the coefficient $C_n$. Eq. (\ref{FokkerPlanckwithexpand}) cannot be solved analytically for the general case
(meaning arbitrary $D_{\mu\mu}$). Even for a specific form of the Fokker-Planck coefficient, a solution cannot be found without using approximations. In the
following we employ the isotropic form for the Fokker-Planck coefficient
\be
D_{\mu\mu} = D \big( 1 - \mu^2 \big).
\label{isoDmumu}
\ee
This form was derived systematically in Shalchi et al. (2009) based on the \textit{second-order quasi-linear theory} developed in Shalchi (2005). It was demonstrated that
the isotropic form should be valid for strong ($\delta B \gg B_0$) and intermediate strong ($\delta B \approx B_0$) turbulence as found in interplanetary and interstellar spaces.
For the isotropic form, Eq. (\ref{FokkerPlanckwithexpand}) becomes
\bdm
& & \sum_{n} \dot{C}_n P_n + i v \mu k_{\parallel} \sum_{n} C_n P_n \nonumber\\
& = & D \sum_{n} C_n \frac{\partial}{\partial \mu} \left[ \left( 1 - \mu^2 \right) \frac{\partial P_n}{\partial \mu} \right]\nonumber\\
& - & \frac{v}{2 L} \left( 1 - \mu^2 \right) \sum_{n} C_n \frac{\partial P_n}{\partial \mu}.
\label{LegendreFokkerPlanck}
\edm
A well-known relation for Legendre polynomials is given by (see, e.g., Abramowitz \& Stegun (1968))
\be
\frac{\partial}{\partial\mu} \left[ \left( 1 - \mu^2 \right) \frac{\partial P_n}{\partial\mu} \right] = -n (n+1) P_n
\ee
where $n$ is integer $n=0,1,2,\dots$. Using this in Eq. (\ref{LegendreFokkerPlanck}) leads to
\bdm
& & \sum_{n} \dot{C}_n P_n + i v \mu k_{\parallel} \sum_{n}C_n P_n \nonumber\\
& = & - D \sum_{n} C_n n(n+1) P_n \nonumber\\
& - & \frac{v}{2 L} \left( 1 - \mu^2 \right) \sum_{n} C_n \frac{\partial P_n}{\partial \mu}.
\label{LegendreFokkerPlanck2}
\edm
To get rid of the remaining explicit $\mu$-dependences we can use the relations (see again Abramowitz \& Stegun (1968))
\be
\mu P_n = \frac{n+1}{2n+1} P_{n+1} + \frac{n}{2n+1} P_{n-1},
\label{recurrence}
\ee
and
\be
\left( 1 - \mu^2 \right) \frac{\partial P_n}{\partial \mu} = n P_{n-1} - n \mu P_{n}.
\ee
Combining the latter two equations yields
\be
\left( 1 - \mu^2 \right) \frac{\partial P_n}{\partial \mu} = n \frac{n+1}{2n+1} \left( P_{n-1} - P_{n+1} \right).
\ee
Therewith Eq. (\ref{LegendreFokkerPlanck2}) becomes
\bdm
& & \sum_{n} \dot{C}_n P_n \nonumber\\
& + & i v k_{\parallel} \sum_{n} C_n \bigg[ \frac{n+1}{2n+1} P_{n+1} + \frac{n}{2n+1} P_{n-1} \bigg] \nonumber\\
& = & - D \sum_{n} C_n n (n+1) P_n\nonumber\\
& + & \frac{v}{2 L} \sum_{n} C_n n \frac{n+1}{2n+1} \left( P_{n+1} - P_{n-1} \right).
\edm
We now multiply this equation by the Legendre polynomial $P_m$, integrate over all $\mu$, and use the orthogonality relation (see, e.g., Abramowitz \& Stegun (1968))
\be
\int_{-1}^{+1} d \mu \; P_n \left( \mu \right) P_m \left( \mu \right) = \frac{2}{2 m + 1} \delta_{nm}
\label{Pnortho}
\ee
 to derive
\bdm
& & \dot{C}_m \nonumber\\
& = & - D m (m+1) C_m \nonumber\\
& - & i v k_{\parallel} \Bigg[ \frac{m}{2m-1} C_{m-1} + \frac{m+1}{2m+3} C_{m+1} \Bigg] \nonumber\\
& + & \frac{v}{2 L} \Bigg[ \frac{m (m - 1)}{2 m - 1} C_{m-1} - \frac{(m + 1)(m + 2)}{2 m + 3} C_{m+1} \Bigg].\nonumber\\
\label{CRelations}
\edm
This leads to an infinite number of coupled ordinary differential equations for the functions $C_m (t)$. We can write those differential equations as the
following linear system:
\be
\dot C = \boldsymbol{M} C
\label{linear_system_cn}
\ee
where we have used
\bdm
C(t)=\left(\begin{array}{c}
C_0(t)    \\[0.2cm]
C_1(t) 	 \\
\vdots 	 \\
\end{array}\right),\qquad \dot C(t)=\left(\begin{array}{c}
\dot C_0(t)      \\[0.2cm]
\dot C_1(t) 	 \\
\vdots 	 \\
\end{array}\right),
\edm
and
\bdm
\boldsymbol{M} = \left(
\begin{array}{cccc}
0   				\quad\quad & M_{01}  		\quad\quad & 0					\quad\quad & \dots		\\[0.2cm]
M_{10}	 		\quad\quad & M_{11}		\quad\quad & M_{12}		\quad\quad & \dots		\\[0.2cm]
0					\quad\quad & M_{21} 		\quad\quad & M_{22}		\quad\quad & \dots		\\[0.2cm]
\vdots			\quad\quad & \vdots			\quad\quad & \vdots			\quad\quad & \ddots							
\end{array}
\right)
\label{eq:M}.
\edm
The elements of the latter matrix are given as the coefficients of $C_n$ and are obtained via Eq.  (\ref{CRelations}). From this it follows that
\bdm
\boldsymbol{M} = \left(
\begin{array}{cccc}
0   						& - \frac{1}{3} i v k_{\parallel}  - \frac{v}{3 L}  	& 0																	& \dots		\\[0.2cm]
- i v k_{\parallel}	& - 2 D															& - \frac{2}{5} i v k_{\parallel} - \frac{3 v}{5 L}		& \dots		\\[0.2cm]
0							& - \frac{2}{3} i v k_{\parallel} + \frac{v}{3 L} 	& - 6 D																& \dots		\\[0.2cm]
\vdots					& \vdots														& \vdots															& \ddots		
\end{array}
\right)\nonumber\\
\label{MatrixwithElements}
\edm
corresponding to an infinite matrix. Eq. (\ref{linear_system_cn}) is a well-known linear system of differential equations with the formal solution
\be
C(t) = e^{\boldsymbol{M} t} C(0)
\label{cn_exact}
\ee
where we have used the matrix exponential. The column vector $C(0)$ is determined using a sharp initial condition of the form
\be
f \big( z, \mu, t=0 \big) = 2 \delta \big( \mu - \mu_0 \big) \delta \big( z \big)
\label{initialfunctionf}
\ee
where $\mu_0$ is the initial pitch-angle cosine. The two Dirac deltas ensure that the particle is initially located at $z=0$ and has the pitch-angle cosine $\mu=\mu_0$.
The factor $2$ was chosen so that
\be
\frac{1}{2} \int_{-1}^{+1} d \mu \int_{-\infty}^{+\infty} d z \; f \big( z, \mu, t=0 \big) = 1
\ee
corresponding to a normalization condition. In order to find the initial conditions for the function $F_{k_{\parallel}} \big( \mu, t \big)$, we need to use the inverse Fourier
transform as given by Eq. (\ref{invFourier}). Combining this with Eq. (\ref{initialfunctionf}) yields
\bdm
F_{k_{\parallel}} \big( \mu, t=0 \big) & = & \frac{1}{2 \pi} \int_{-\infty}^{+\infty} d z \; f \big( z, \mu, t=0 \big) e^{- i k_{\parallel} z} \nonumber\\
& = & \frac{1}{\pi} \delta \big( \mu - \mu_0 \big) \int_{-\infty}^{+\infty} d z \; \delta \big( z \big) e^{- i k_{\parallel} z} \nonumber\\
& = & \frac{1}{\pi} \delta \big( \mu - \mu_0 \big).
\edm
In combination with expansion (\ref{expandF}) this gives us
\be
\sum_{n} C_n (t=0) P_n \big( \mu \big) = \frac{1}{\pi} \delta \big( \mu - \mu_0 \big).
\label{InitialCondition}
\ee
Multiplying this equation by $P_m$ and invoking Eq.  (\ref{Pnortho}) results in
\be
C_m \big( t = 0 \big) = \frac{2 m + 1}{2 \pi} P_m \big( \mu_0 \big)
\label{InitialCoeffRelation}
\ee
Orthogonality of the Legendre polynomials also allows us to derive the following expression for the $\mu$-integrated solution:
\bdm
M (z, t) & = & \frac12 \int_{-1}^{+1} d \mu \; f (z, \mu, t) \nonumber\\
& = & \frac12 \int_{-\infty}^{+\infty} d k_{\parallel} \; e^{i k_\parallel z} \int_{-1}^{+1} d \mu \; F_{k_{\parallel}} (\mu, t)\nonumber\\
& = & \int_{-\infty}^{+\infty} d k_{\parallel} \; C_0 (k_\parallel, t) e^{i k_\parallel z}.
\label{sub_mu_integral}
\edm
Unfortunately, Eq. (\ref{cn_exact}) cannot be solved analytically and exactly. Therefore, we need to truncate the expansion provided by Eq. (\ref{expandF}). That is, set
\be
C_n = 0 ,\qquad n\geq N
\ee
for some $N$,  which corresponds to what we call the \textit{$N$-dimensional subspace approximation}. Evidently, the larger we take $N$ to be, the better our approximation is.
Shalchi (2024) concluded that when dealing with Eq. (\ref{FPeq}) without the focusing term, $N=10$ corresponds to a fast and accurate approximation. In Sect. 5, we extend
this idea with focusing included. 

Alternatively, we can solve Eq. (\ref{linear_system_cn}) analytically if $N$ is small enough. This can be done by determining the eigenvalues of the $N \times N$ matrix $\textbf{M}$.
In Lasuik \& Shalchi (2019), it was concluded that $N=2$ results in an accurate but simple approximation. However, the latter work is based on the transport equation without
focusing term. Section 3 discusses a similar approach but with the focusing term included.

\section{The Two-Dimensional Subspace Approximation}\label{2dsection}

In the case of the two-dimensional subspace approximation, the expansion provided by Eq. (\ref{expandF}) is simplified to
\be
F_{k_{\parallel}} (\mu, t) = C_0(t) + \mu C_1(t).
\label{expandF2D}
\ee
Additionally, Eq. (\ref{linear_system_cn}) becomes
\bdm
\left(
\begin{array}{c}
\dot{C}_0      \\[0.2cm]
\dot{C}_1      \\
\end{array}
\right) = \left(
\begin{array}{ccc}
0   							\quad & - \frac{1}{3} i v k_{\parallel} - \frac{v}{3 L} \\[0.2cm]
- i v k_{\parallel}		\quad & - 2 D
\end{array}
\right) \left(
\begin{array}{c}
C_0   \\[0.2cm]
C_1 	 \\
\end{array}
\right).\nonumber\\
\label{2matrix}
\edm
The \textit{ansatz}
\be
C_0 \big( t \big) = b_0 e^{\omega t} \quad \textnormal{and} \quad C_1 \big( t \big) = b_1 e^{\omega t}
\label{ansatz}
\ee
leads to the matrix equation
\bdm
\omega \left(
\begin{array}{c}
b_0      \\[0.2cm]
b_1      \\
\end{array}
\right) = \left(
\begin{array}{ccc}
0   							\quad & - \frac{1}{3} i v k_{\parallel} - \frac{v}{3 L} \\[0.2cm]
- i v k_{\parallel}		\quad & - 2 D
\end{array}
\right) \left(
\begin{array}{c}
b_0   \\[0.2cm]
b_1 	 \\
\end{array}
\right)
\edm
corresponding to a simple eigenvalue problem. Alternatively, this can be written as
\bdm
\left(
\begin{array}{ccc}
- \omega   				\quad & - \frac{1}{3} i v k_{\parallel} - \frac{v}{3 L} \\[0.2cm]
- i v k_{\parallel}		\quad & - 2 D - \omega
\end{array}
\right) \left(
\begin{array}{c}
b_0      \\[0.2cm]
b_1 	 \\
\end{array}
\right) = 0.
\edm
Non-trivial solutions of the latter equation are obtained by setting the determinant of this $2 \times 2$ matrix equal to zero. We find
\be
\omega^2 + 2 D \omega + \frac{1}{3} v^2 k_{\parallel}^2 - i \frac{v^2 k_{\parallel}}{3 L} = 0.
\ee
This quadratic equation has the two solutions
\be
\omega_{\pm} = - D \pm \sqrt{D^2 - \frac{1}{3} v^2 k_{\parallel}^2 + i \frac{v^2 k_{\parallel}}{3 L}}.
\label{omegapm}
\ee
These two values for $\omega$ can be used in the \textit{ansatz} given by Eq. (\ref{ansatz}). The general solution for the function $C_0 (t)$ is, therefore, obtained via the superposition
\be
C_0 \big( t \big) = b_{+} e^{\omega_{+} t} + b_{-} e^{\omega_{-} t}.
\label{superpos}
\ee
By combining Eqs.  (\ref{2matrix}) and (\ref{superpos}),  we also obtain
\be
C_1 \big( t \big) = \frac{3 i}{v k_{\parallel} - i v /  L} \left( \omega_{+} b_{+} e^{\omega_{+} t} + \omega_{-} b_{-} e^{\omega_{-} t} \right).
\label{SolforC1}
\ee
Eq.  (\ref{InitialCoeffRelation}) states that
\be
C_0 (0) = \frac1{2\pi} \quad\textnormal{and}\quad C_1 (0) =\frac{3}{2 \pi} \mu_0.
\label{initial_c0,c1}
\ee
Together, Eqs. (\ref{superpos})-(\ref{initial_c0,c1}) imply the linear system for $b_{+}$ and $b_{-}$:
\be
b_+ + b_- =\frac{1}{2 \pi}
\label{Initialb1}
\ee
and
\be
\frac{3 i}{v k_{\parallel} - i v /  L} \left( \omega_{+} b_{+} + \omega_{-} b_{-} \right) = \frac{3 \mu_0}{2 \pi}.
\label{Initialb2}
\ee
Eqs. (\ref{Initialb1}) and (\ref{Initialb2}) can be combined to determine the parameters we are looking for. After some straightforward algebra we find
\bdm
b_{+} & = & - \frac{1}{2 \pi} \frac{\omega_{-} + i v \mu_0 k_{\parallel} + v \mu_0 / L}{\omega_{+} - \omega_{-}},\nonumber\\
b_{-} & = & \frac{1}{2 \pi} \frac{\omega_{+} + i v \mu_0 k_{\parallel} + v \mu_0 /  L}{\omega_{+} - \omega_{-}}.
\label{thetwobs}
\edm
These formulas can be used in Eqs. (\ref{superpos}) and (\ref{SolforC1}) and our coefficients $C_0 (t)$ and $C_1 (t)$ are completely determined. Our
solution for the Fourier-transformed distribution function as provided by Eq. (\ref{expandF2D}) can, therefore, be written as
\bdm
F_{k_{\parallel}}\big(\mu, t \big) & = & \left[ 1 + \frac{3 i \mu \omega_{+}}{v k_{\parallel} - i v / L} \right] b_{+} e^{\omega_{+} t}\nonumber\\
& + & \left[ 1 + \frac{3 i \mu \omega_{-}}{v k_{\parallel} - i v / L} \right] b_{-} e^{\omega_{-} t}
\label{finalF}
\edm
where the coefficients $b_{\pm}$ are given by Eq. (\ref{thetwobs}) and the coefficients $\omega_{\pm}$ are provided by Eq. (\ref{omegapm}). Of course, our solution
is based on the two-dimensional subspace approximation and is, therefore, not an exact solution of Eq. (\ref{FourierFPeq}).

One must be cautious when the two-dimensional subspace approximation is used as there are values of $v/(DL)$ which can cause either of the eigenvalues to have a positive
real component for large wave numbers. This would lead to an infinitely large $F_{k_\parallel}$. To demonstrate this, we consider Eq. (\ref{omegapm}) and explore the limit
$k_{\parallel} \rightarrow \infty$. We assume that $L>0$ and first notice that 
\bdm
&&\Re  \left\{ \sqrt{D^2 - \frac{1}{3} v^2 k_{\parallel}^2 + i \frac{v^2 k_{\parallel}}{3 L}} \right\}\nonumber\\
&&=\left[\left( D^2 - \frac{1}{3} v^2 k_{\parallel}^2\right)^2+ \frac{v^4 k_{\parallel}^2}{9 L^2}\right]^{1/4}
\nonumber\\
&&\times\cos\left[\frac12\tan^{-1}\left(\frac{\frac{v^2 k_{\parallel}}{3 L}}{D^2 - \frac{1}{3} v^2 k_{\parallel}^2} \right)+\frac12\pi\right]
\edm
where $\Re$ denotes the real component. One can easily see here that in the limit $k_{\parallel}\to\infty$, the indeterminate form $\infty\cdot 0$ arises. Evaluating the limit is
then a matter of applying l'H\^{o}pital's rule. After a tedious yet straightforward calculation, we derive
\be
\lim_{k_{\parallel}\to\infty} \Re \left\{ \sqrt{D^2 - \frac{1}{3} v^2 k_{\parallel}^2 + i \frac{v^2 k_{\parallel}}{3 L}} \right\} = \frac v{2\sqrt3 L}
\ee 
As a result,  we can see
\be
\lim_{k_{\parallel}\to\infty} \Re \left\{ \omega_+ \right\} = -D+ \frac v{2\sqrt3 L}
\ee
which is positive if
\be
\frac{v}{DL} > 2 \sqrt{3}.
\label{pos_eig}
\ee
Similarly, if $L<0$, we have that $\omega_-$ has a positive real component for large wave numbers if
\be
-\frac{v}{DL} > 2 \sqrt3.
\ee
Therefore, there is a limit for the possible values of the parameter $v/(DL)$.

It was demonstrated in Shalchi (2024) that in the absence of focusing, the two-dimensional subspace approximation is equivalent to using a telegraph equation.
Using a telegraph equation description of energetic particle transport goes back to the early work of Fisk \& Axford (1969) in which the authors focused on early times
in solar flare events, where the particle inertia needs to be accounted for before a diffusive equilibrium sets in.

\section{Expectation Values}

By using the results derived in the previous section, we can calculate several important quantities. First we need to define the meaning of the ensemble average in particle transport
theory. Since the particle distribution function derived above is a function of $\mu_0$, $\mu$, and $z$, we understand the ensemble average of the quantity $A (z,\mu,t)$ as
\bdm
\big< A \big> & = & \frac{1}{4} \int_{-1}^{+1} d \mu \; \int_{-1}^{+1} d \mu_0 \nonumber\\
& \times & \int_{-\infty}^{+\infty} d z \; A \left( z, \mu, t \right) f \big( z, \mu, t \big).
\label{defaverageoperator}
\edm
Sometimes one can drop one of the integrals therein. This is in particular the case if a pitch-angle dependent solution is needed. In the following subsections we consider several
examples for the quantity $A$ such as the characteristic function, the velocity correlation function, and moments of the form $\langle z^n \rangle$. To know the analytical form
of such quantities is essential to develop analytical theories for perpendicular transport (see, e.g., Matthaeus et al. (2003), Shalchi (2010), Shalchi (2020), and Shalchi (2021).

\subsection{The Characteristic Function}\label{char_function}

As a first application, we compute the characteristic function defined via
\bdm
\left< e^{- i k_{\parallel} z} \right>
& = & \frac{1}{4} \int_{-1}^{+1} d \mu \int_{-1}^{+1} d \mu_0 \nonumber\\
& \times & \int_{-\infty}^{+\infty} d z \; f \big( z, \mu, t \big) e^{- i k_{\parallel} z}.
\label{eq:characteristic}
\edm
Therein we can use Eq. (\ref{invFourier}) to find
\be
\left< e^{- i k_{\parallel} z} \right>
= \frac{\pi}{2} \int_{-1}^{+1} d \mu \int_{-1}^{+1} d \mu_0 \; F_{k_\parallel} \big(\mu, t \big)
\ee
meaning that the desired characteristic function is nothing else than the $\mu_0$- and $\mu$-averaged function $F_{k_\parallel}$. If we average Eq. (\ref{expandF}) over all $\mu$,
we find
\be
\frac{1}{2} \int_{-1}^{+1} d \mu \;  F_{k_\parallel} \big(\mu, t \big) = C_0 \big( t \big)\label{eq:Fint}
\ee
where we have used orthogonality of the Legendre polynomials as given by Eq. (\ref{Pnortho}). Using the result provided by Eq. (\ref{superpos}) therein yields
\be
\left< e^{- i k_{\parallel} z} \right> = \pi \int_{-1}^{+1} d \mu_0 \; \left[ b_{+} e^{\omega_{+} t} + b_{-} e^{\omega_{-} t} \right].
\ee
After employing Eq. (\ref{thetwobs}) we can solve the remaining integral so that
\be
\left< e^{- i k_{\parallel} z} \right> = \frac{\omega_{+}}{\omega_{+} - \omega_{-}} e^{\omega_{-} t} - \frac{\omega_{-}}{\omega_{+} - \omega_{-}} e^{\omega_{+} t}
\label{charfunc}
\ee
where the parameters $\omega_{+}$ and $\omega_{-}$ are given by Eq. (\ref{omegapm}). Taking the complex conjugate of Eq. (\ref{charfunc}) yields
\be
\left< e^{+ i k_{\parallel} z} \right> = \frac{\omega_{+}^{*}}{\omega_{+}^{*} - \omega_{-}^{*}} e^{\omega_{-}^{*} t}
- \frac{\omega_{-}^{*}}{\omega_{+}^{*} - \omega_{-}^{*}} e^{\omega_{+}^{*} t}.
\label{charfunccc}
\ee
For the case that $\omega_{+}$ and $\omega_{-}$ are real, we can easily see that
\be
\left< e^{+ i k_{\parallel} z} \right> = \left< e^{- i k_{\parallel} z} \right>.
\label{ccisthesame}
\ee
For complex $\omega_{\pm}$ it follows from Eq. (\ref{omegapm}) that $\omega_{\pm}^{*} = \omega_{\mp}$. Using this in Eq. (\ref{charfunccc}) gives us the
right-hand-side of Eq. (\ref{charfunc}) and, thus, we conclude that Eq. (\ref{ccisthesame}) is valid in general.

The characteristic function discussed above is directly related to the $\mu_0$- and $\mu$-averaged distribution function
\be
M \left( z, t \right) = \frac{1}{2 \pi} \int_{-\infty}^{+\infty} d k_{\parallel} \; \langle e^{\pm i k_{\parallel} z} \rangle e^{i k_{\parallel} z}.
\label{FourierChar}
\ee
Therein we can use results for the characteristic function such as the one given by Eq. (\ref{charfunc}).

In the following we consider special cases with the aim to simplify Eq. (\ref{charfunc}).

\subsubsection{Small Wave Numbers}

It is convenient to use the short notation
\be
\sqrt{\rule{0pt}{2ex}\dots} := \sqrt{ D^2 - \frac{1}{3} v^2 k_{\parallel}^2 + i \frac{v^2 k_{\parallel}}{3 L}}
\label{expandsqrt}
\ee
in Eq. (\ref{omegapm}). First we assume that
\be
\left|-v^2k_\parallel^2+i\frac{v^2k_\parallel}L\right| \ll 3D^2.
\label{smallkpara}
\ee
After Taylor-expanding the square root given by Eq. (\ref{expandsqrt}), we find up to $k_{\parallel}^2$ the approximation
\be
\sqrt{\rule{0pt}{2ex}\dots} \approx D + i \frac{v^2}{6 D L} k_{\parallel} + \left( \frac{v^4}{72 L^2 D^3} - \frac{v^2}{6 D} \right) k_{\parallel}^2.
\ee
For isotropic pitch-angle scattering and without adiabatic focusing, the parallel spatial diffusion coefficient is given by (see, e.g., Shalchi (2006))
\be
\kappa_{\parallel} = \frac{v^2}{6 D}.
\label{eq:kparallel}
\ee
Therewith, our result can be written as
\be
\sqrt{\rule{0pt}{2ex}\dots} \approx D + i \frac{\kappa_{\parallel}}{L} k_{\parallel}
- \kappa_{\parallel} \left( 1 - \frac{v^2}{12 L^2 D^2} \right) k_{\parallel}^2.
\ee
Using this expansion in Eq. (\ref{omegapm}) yields
\be
\omega_{+} \approx i \frac{\kappa_{\parallel}}{L} k_{\parallel} - \kappa_{\parallel} \left( 1 - \frac{v^2}{12 L^2 D^2} \right) k_{\parallel}^2
\label{strongomegap}
\ee
as well as
\be
\omega_{-} \approx - \frac{v}{\lambda_{\parallel}} - i \frac{\kappa_{\parallel}}{L} k_{\parallel}+ \kappa_{\parallel} \left( 1 - \frac{v^2}{12 L^2 D^2} \right) k_{\parallel}^2
\label{omegaminuswithfocus}
\ee
where we have used the parallel mean free path without focusing effect (see again Shalchi (2006))
\be
\lambda_{\parallel} = \frac{v}{2 D}.
\label{MFPnofocusing}
\ee
In general, the parallel mean free path is related to the parallel spatial diffusion coefficient via $\lambda_{\parallel} = 3 \kappa_{\parallel} / v$. The above expansion is valid as long as
Eq. (\ref{smallkpara}) holds. Therefore, we can further simplify the result given by Eq. (\ref{omegaminuswithfocus}) by only keeping the lowest order terms for both the real and imaginary
components to obtain
\be
\omega_{-} \approx - \frac{v}{\lambda_{\parallel}} - i \frac{\kappa_{\parallel}}{L} k_{\parallel}.
\label{strongomegam}
\ee
Then, on the other hand, we can write Eq. (\ref{strongomegap}) as
\be
\omega_{+} \approx - \bar{\kappa}_{\parallel} k_{\parallel}^2 + i \frac{\kappa_{\parallel}}{L} k_{\parallel}
\label{strongomegap2}
\ee
where we have used the parallel diffusion coefficient with focusing
\bdm
\bar{\kappa}_{\parallel} & = & \kappa_{\parallel} \left( 1 - \frac{v^2}{12 L^2 D^2} \right) \nonumber\\
& = & \kappa_{\parallel} \left( 1 - \frac{1}{3} \frac{\lambda_{\parallel}^2}{L^2} \right).
\label{defkappabar}
\edm
Note, this reduction of the parallel diffusion coefficient due to focusing was found before (see, e.g., Shalchi (2011b) and Shalchi \& Danos (2013)). Quantitatively, the
result derived here is slightly different compared to previous results. One reason is that it depends on how a diffusion coefficient is defined. It was demonstrated in
Danos et al. (2013) that diffusion coefficients defined via mean square displacements and diffusion coefficients defined via a time-integral over the velocity correlation
functions are not the same if there is focusing. Furthermore, Eq. (\ref{defkappabar}) is based on the two-dimensional subspace approximation and is, therefore, not
an exact result.

In the limit considered here the parameters $\omega_{-}$ and $\omega_{+}$ are given by Eqs. (\ref{strongomegam}) and (\ref{strongomegap2}), respectively.
Using this in Eq. (\ref{charfunc}) gives us
\bdm
\left< e^{- i k_{\parallel} z} \right>
& = & \frac{i \frac{\kappa_{\parallel}}{L} k_{\parallel} - \bar{\kappa}_{\parallel} k_{\parallel}^2}{2 \sqrt{\rule{0pt}{2ex}\dots}}
e^{- v t / \lambda_{\parallel} - i \kappa_{\parallel} k_{\parallel} t / L}\nonumber\\
& + & \frac{\frac{v}{\lambda_{\parallel}} + i \frac{\kappa_{\parallel}}{L} k_{\parallel}}{2 \sqrt{\rule{0pt}{2ex}\dots}}
e^{i \kappa_{\parallel} k_{\parallel} t / L - \bar{\kappa}_{\parallel} k_{\parallel}^2 t}
\edm
where we have also employed Eq. (\ref{ccisthesame}). Since we have used already the condition given by Eq. (\ref{smallkpara}), we can further simplify our result by 
\be
\left< e^{- i k_{\parallel} z} \right> = e^{i \kappa_{\parallel} k_{\parallel} t / L - \bar{\kappa}_{\parallel} k_{\parallel}^2 t}.
\label{chardiff}
\ee
In Sect. (\ref{sec:early_late}), a more detailed discussion of this result can be found.

\subsubsection{Large Wave Numbers}

In the previous paragraph we considered the case of small wave numbers. However, we can also approximate Eq. (\ref{charfunc}) by using the limit
\be
v^2k_\parallel^2 \gg \left| 3 D^2 + i \frac{v^2k_\parallel} L \right|.
\label{largekpara}
\ee
In this case Eq. (\ref{omegapm}) becomes
\bdm
\omega_{\pm}&=&-D\pm\left(i\frac{v|k_\parallel|}{\sqrt3}+ \frac{v\text{sgn}(k_\parallel)}{2\sqrt3 L}\right.\nonumber\\
&-& \left.i\frac{\sqrt3 D^2}{2v|k_\parallel|} +i\frac{v}{8\sqrt3 L^2|k_\parallel|}\right)\nonumber\\
&\approx& -D\pm\left(i\frac{v|k_\parallel|}{\sqrt3}+ \frac{v\text{sgn}(k_\parallel)}{2\sqrt3 L}\right)
\edm
where $\text{sgn}(x)=1$ if $x\geq 0$ and $\text{sgn}(x)=-1$ if $x<0$. Furthermore, it follows from Eq. (\ref{omegapm}) that
\bdm
2\sqrt{\rule{0pt}{2ex}\dots}&=&\frac{v\text{sgn}(k_\parallel)}{\sqrt3 L}+i\left(\frac2{\sqrt3}v|k_\parallel|-\frac{\sqrt3 D^2}{v|k_\parallel| }\right.\nonumber\\
&+&\left. \frac{v}{4\sqrt3 L^2|k_\parallel|}\right)\nonumber\\
&\approx& \frac{v\text{sgn}(k_\parallel)}{\sqrt3 L}+i\frac2{\sqrt3}v|k_\parallel|.
\edm
Using this in Eq. (\ref{charfunc}) yields
\bdm
& & \left< e^{- i k_{\parallel} z} \right>\nonumber\\
&=&\frac{2k_\parallel L}{1 + 2 i k_\parallel L}\nonumber\\
&\times&\Bigg[\left(-\frac{\sqrt3}2\frac{D}{v|k_\parallel|}+\frac1{4k_\parallel L}+i\frac12\right)e^{-\left( \frac{v\text{sgn}(k_\parallel)}{2\sqrt3 L}+i\frac{v|k_\parallel|}{\sqrt3}\right)t} \nonumber\\
&+&\left(\frac{\sqrt3}2\frac{D}{v|k_\parallel|}+\frac1{4k_\parallel L}+i\frac12\right)e^{\left(\frac{v\text{sgn}(k_\parallel)}{2\sqrt3 L}+i\frac{v|k_\parallel|}{\sqrt3}\right)t}\Bigg] e^{-Dt}.\nonumber\\
\edm
If we additionally assume that $L, |k_\parallel|\to\infty$, we can further simplify this result to obtain
\bdm
\left< e^{- i k_{\parallel} z} \right> & = & \frac{1}{2} \left[ e^{i\frac{vk_\parallel}{\sqrt 3}t} + e^{-i\frac{vk_\parallel}{\sqrt 3}t} \right] e^{- D t}\nonumber\\
& = & \cos\left(\frac{vk_\parallel}{\sqrt3}t\right)e^{-Dt}.
\label{eq:large}
\edm
The latter result is easy to understand. It corresponds to the unperturbed orbit but damped due to pitch-angle scattering. Note, this result was also derived in Lasuik \& Shalchi (2019).
In the considered order, Eq. (\ref{eq:large}) does not depend on the focusing effect.

\subsubsection{Early and Late Times}\label{sec:early_late}

In general the characteristic function is given by Eq. (\ref{charfunc}). Combining this formula with Eq. (\ref{FourierChar}) yields
\bdm
& & M \big( z, t \big) \nonumber\\
& = & \frac{1}{2 \pi} \int_{-\infty}^{+\infty} d k_{\parallel} \; \bigg[ \frac{\omega_{+}}{\omega_{+} - \omega_{-}} e^{\omega_{-} t}
- \frac{\omega_{-}}{\omega_{+} - \omega_{-}} e^{\omega_{+} t} \bigg] e^{i k_{\parallel} z}.\nonumber\\
\label{genM}
\edm
For $t = 0$ this becomes
\bdm
& & M \big( z, t=0 \big) \nonumber\\
& = & \frac{1}{2 \pi} \int_{-\infty}^{+\infty} d k_{\parallel} \; \bigg[ \frac{\omega_{+}}{\omega_{+} - \omega_{-}}
- \frac{\omega_{-}}{\omega_{+} - \omega_{-}} \bigg] e^{i k_{\parallel} z} \nonumber\\
& = & \frac{1}{2 \pi} \int_{-\infty}^{+\infty} d k_{\parallel} \; e^{i k_{\parallel} z} \nonumber\\
& = & \delta \big( z \big)
\label{Mt0}
\edm
where in the last step we have used (see, e.g., Zwillinger (2012))
\be
\int_{-\infty}^{+\infty} d z \; e^{i (k_{\parallel}^{\prime}-k_{\parallel}) z} = 2 \pi \delta \big( k_{\parallel}^{\prime}-k_{\parallel} \big).
\label{zwillinger}
\ee
Eq. (\ref{Mt0}) corresponds to the pitch-angle averaged initial distribution.

For $t \rightarrow \infty$, on the other hand, the main contribution to the integral in Eq. (\ref{genM}) comes from the smallest possible values of $\omega_{\pm}$
due to the exponential in the aforementioned equation. It follows from Eq. (\ref{omegapm}) that this happens for the smallest possible wave number $k_{\parallel}$.
Therefore, we find with the help of Eq. (\ref{chardiff}) that
\bdm
& & M \big( z, t \rightarrow \infty \big) \nonumber\\
& = & \frac{1}{2 \pi} \int_{-\infty}^{+\infty} d k_{\parallel} \; e^{i \kappa_{\parallel} k_{\parallel} t / L - \bar{\kappa}_{\parallel} k_{\parallel}^2 t} e^{i k_{\parallel} z}\nonumber\\
& = & \frac{1}{\sqrt{4 \pi \bar{\kappa}_{\parallel} t}} e^{- (z + \kappa_{\parallel} t / L)^2 / (4 \bar{\kappa}_{\parallel} t)}
\label{GaussianM}
\edm
where the parameter $\bar{\kappa}_{\parallel}$ is given by Eq. (\ref{defkappabar}). Eq. (\ref{GaussianM}) corresponds to a Fourier-transformed Gaussian distribution
with non-vanishing mean. From Eq. (\ref{GaussianM}), we can easily read off the mean position
\be
\big< z \big> = - \frac{\kappa_{\parallel} t}{L}
\label{eq:mean}
\ee
and the mean square displacement
\be
\big< z^2 \big> - \big< z \big>^2 = 2 \bar{\kappa}_{\parallel} t.
\label{zMSD}
\ee
Eq. (\ref{eq:mean}) is a well-known result in the theory of adiabatic focusing (see Shalchi (2009b)). The above findings corresponds to the solution of a diffusion-convection equation.
The mean of the Gaussian distribution moves with the speed $v_c = - \kappa_{\parallel} / L$ whereas the width is given by Eq. (\ref{zMSD}). The parallel diffusion coefficient therein is altered
by adiabatic focusing as given by Eq. (\ref{defkappabar}).

\subsection{The Velocity Correlation Function}

We can also use the formulation discussed so far to determine the parallel velocity correlation function. This can be done via
\bdm
& & V_{zz} \big( t \big) \nonumber\\
& = & v^2 \Big< \mu_0 \mu \Big> \nonumber\\
& = & \frac{v^2}{4} \int_{-1}^{+1} d \mu_0 \int_{-1}^{+1} d \mu \; \int_{-\infty}^{+\infty} d z \; \mu_0 \mu f \big( z, \mu, t \big).\nonumber\\
\label{eq:velocitycorrelation}
\edm
Therein we can use
\bdm
& & \int_{-\infty}^{+\infty} d z \; f \big( z, \mu, t \big)\nonumber\\
& = & \int_{-\infty}^{+\infty} d k_{\parallel} \; F_{k_{\parallel}} \big(\mu, t \big) \int_{-\infty}^{+\infty} d z \; e^{i k_{\parallel} z} \nonumber\\
& = & 2 \pi F_0 \big(\mu, t \big)
\edm
where we have employed again Eq. (\ref{zwillinger}). Therefore, the velocity correlation functions becomes
\bdm
V_{zz} \big( t \big)
& = & \frac{\pi}{2} v^2 \int_{-1}^{+1} d \mu_0 \int_{-1}^{+1} d \mu \; \mu_0 \mu F_0 \big(\mu, t \big) \nonumber\\
& = & \frac{\pi}{2} v^2 \int_{-1}^{+1} d \mu_0 \; \mu_0 \int_{-1}^{+1} d \mu \; \mu^2 C_1 \big( k_{\parallel}=0, t \big) \nonumber\\
& = & \pi \frac{v^2}{3} \int_{-1}^{+1} d \mu_0 \; \mu_0 C_1 \big( k_{\parallel}=0, t \big)\label{eq:velocitycorrsimp}
\edm
where in the second step, we have again used the orthogonal relation given by Eq. (\ref{Pnortho}). Using therein Eq. (\ref{SolforC1}) with (\ref{thetwobs}) yields
\be
V_{zz} \big( t \big) = \frac{v^2}{3} \bigg[ \frac{\omega_{+}}{\omega_{+} - \omega_{-}} e^{\omega_{+} t}
- \frac{\omega_{-}}{\omega_{+} - \omega_{-}} e^{\omega_{-} t} \bigg]_{k_{\parallel} = 0}.
\ee
For $k_{\parallel} = 0$ we have
\be
\omega_+ (k_\parallel=0) = 0 \quad\textnormal{and}\quad \omega_-(k_\parallel=0) = - 2 D.
\label{omegasfork0}
\ee
Therewith we finally find for the velocity correlation function
\be
\left< V_z (t) V_z (0) \right> = V_{zz} \big( t \big) = \frac{v^2}{3} e^{- 2 D t} = \frac{v^2}{3} e^{-v t / \lambda_{\parallel}}
\ee
where we have used the parallel mean free path without focusing effect as given by Eq. (\ref{MFPnofocusing}). As demonstrated, the velocity correlation function is an exponential
function for the isotropic pitch-angle Fokker-Planck coefficient given by Eq. (\ref{isoDmumu}). The characteristic length for the decay of the obtained exponential is the parallel mean
free paths. For other forms of the pitch-angle Fokker-Planck coefficient, different velocity correlation functions can be obtained (see Shalchi (2011a)). With velocity correlation functions
there is an associated diffusion coefficient which is computed via the \textit{TGK (Taylor-Green-Kubo) formulation} (see Taylor (1922), Green (1951), and Kubo (1957))
\bdm
\kappa_{\parallel}^{TGK} & = & \int_{0}^{\infty} d t \; \left< V_z (t) V_z (0) \right> \nonumber\\
& = & \frac{v^2}{3} \int_{0}^{\infty} d t \; e^{- 2 D t} \nonumber\\
& = & \frac{v^2}{6 D}.
\label{tgk}
\edm
The latter result agrees with Eq. (\ref{eq:kparallel}). We conclude that within the two-dimensional subspace approximation,  adiabatic focusing has no effect on velocity
correlation functions and the associated parallel diffusion coefficient $\kappa_{\parallel}^{TGK}$. However, in Appendix \ref{sec:3dim} we show that it does have an effect
if we instead use the three-dimensional subspace approximation instead.

\subsection{The Expectation Value $\left<\mu \right>$}\label{sec:2d_mu}

In the following we compute the mean pitch-angle cosine $\left<\mu \right>$ by employing the two-dimensional subspace approximation. This quantity is obtained via
\be
\left<\mu \right> = \frac12\int_{-1}^{+1} d \mu \; \int_{-\infty}^\infty dz \; \mu f(z,\mu,t).
\ee
With Eq. (\ref{FourierTransform}), this becomes
\bdm
\left<\mu\right> & = & \frac12\int_{-1}^{+1}  d\mu \mu \int_{-\infty}^\infty dk_\parallel \; F_{k_\parallel} (\mu,t) \int_{-\infty}^{+\infty} d z \; e^{i k_\parallel z}\nonumber\\
& = & \pi \int_{-1}^{+1} d\mu \mu \int_{-\infty}^{+\infty} dk_\parallel \; F_{k_\parallel}(\mu,t)\delta(k_\parallel)\nonumber\\
& = & \pi \int_{-1}^{+1} d\mu \; \mu  F_0(\mu,t)
\edm
where we have used again Eq. (\ref{zwillinger}). Combining this with Eq.  (\ref{Pnortho}) yields
\be
\left<\mu\right> = \frac23\pi C_1(k_\parallel=0,t).
\label{eq:exmu}
\ee
For $k_\parallel = 0$ we can employ Eq. (\ref{omegasfork0}). Using this and Eq. (\ref{SolforC1}) in (\ref{eq:exmu}) provides
\be
\left< \mu \right> = \mu_0 e^{-2Dt}.
\label{eq:expec_mu}
\ee
Again, we see that within the two-dimensional subspace approximation, adiabatic focusing has no effect on this result. However, this is not the case for the three-dimensional
subspace approximation, as shown in Appendix \ref{sec:3dim}.

\subsection{The Moments $\left< z^n \right>$}

The moments $\left< z^n \right>$ are obtained via
\bdm
\left<z^n\right> & = & \frac12 \int_{-1}^{+1} d \mu  \int_{-\infty}^{+\infty} dz \; z^n f (z,\mu,t)\nonumber\\
& = & \frac12 \int_{-1}^{+1} d\mu \; \int_{-\infty}^{+\infty} d k_\parallel \; F_{k_\parallel} (\mu,t)\nonumber\\
& \times & \int_{-\infty}^{+\infty} dz \; z^n e^{ik_\parallel z}\nonumber\\
\label{zn}
\edm
where we have used Eq.  (\ref{FourierTransform}). Using
\be
z^n e^{i k z} = (-i)^n \frac{\partial^n}{\partial k_\parallel^n} e^{i k_\parallel z}
\ee
in Eq. (\ref{zn}) yields
\bdm
\left< z^n \right> 
& = & (-i)^n \frac12 \int_{-1}^{+1} d\mu \int_{-\infty}^{+\infty} d k_\parallel \; F_{k_\parallel} (\mu,t)\nonumber\\
& \times & \int_{-\infty}^{+\infty} dz \; \frac{\partial^n}{\partial k_\parallel^n} e^{i k_\parallel z}.
\label{zn2}
\edm
Integrating by parts $n$-times yields
\bdm
\left< z^n \right> & = & i^n \frac{1}{2} \int_{-1}^{+1} d \mu \int_{-\infty}^{+\infty} d k_\parallel \; \frac{\partial^n}{\partial k_\parallel^n} F_{k_\parallel} (\mu,t) \nonumber\\
& \times & \int_{-\infty}^{+\infty} d z \; e^{i k_\parallel z}.
\edm
The $z$-integral can be evaluated with the help of Eq. (\ref{zwillinger}) giving us
\be
\left< z^n \right> = i^n \pi \int_{-1}^{+1} d \mu \int_{-\infty}^\infty d k_\parallel \; \delta (k_\parallel) \frac{\partial^n}{\partial k_\parallel^n} F_{k_\parallel} (\mu,t).
\ee
The integral over the Dirac delta can be evaluated and we find
\be
\left< z^n \right> = i^n \pi \left[ \frac{\partial^n}{\partial k_\parallel^n} \int_{-1}^{+1} d\mu \; F_{k_\parallel} (\mu,t) \right]_{k_\parallel=0}.
\label{zn_simplified}
\ee
To evaluate this further we use Eq. (\ref{expandF2D}) and employ the orthogonality relation provided by Eq. (\ref{Pnortho}) to obtain
\be
\left< z^n \right> = i^n 2 \pi \left[ \frac{\partial^n}{\partial k_\parallel^n} C_0 \left( k_\parallel, t \right) \right]_{k_\parallel=0}.
\label{eq:exzn}
\ee
As a first example we consider $n=1$. Using Eq. (\ref{superpos}) in (\ref{eq:exzn}) yields
\bdm
\left< z \right> & = & i 2 \pi \Bigg[ e^{\omega_+t} \frac{d}{dk_\parallel} b_+ + b_+ e^{\omega_+t} t \frac{d}{dk_\parallel} \omega_+\nonumber\\
& \qquad & \qquad + e^{\omega_-t}\frac{d}{dk_\parallel} b_- + b_- e^{\omega_-t} t \frac{d}{dk_\parallel} \omega_- \Bigg]_{k_\parallel=0}.\nonumber\\
\edm
It follows from Eq.  (\ref{omegapm}) that
\be
\frac{d\omega_{\pm}}{d k_\parallel}\Big\vert_{k_\parallel=0} = \pm i\frac{v^2}{6DL},
\label{domegadk}
\ee
and from Eq. (\ref{thetwobs}) we get
\be
\frac{d  b_{\pm}}{d k_\parallel}\Big\vert_{k_\parallel=0} =\mp\frac{i}{8\pi D^2}\left[\frac13\frac{v^2}{L}+2Dv\mu_0-\frac13\frac{v^3\mu_0}{DL^2}\right].
\label{dbdk}
\ee
We also use Eq. (\ref{omegasfork0}) as well as
\bdm
b_+ (k_\parallel=0) & = & \frac1{2\pi} \frac{2 D L - v \mu_0}{2 D L},\nonumber\\
b_- (k_\parallel=0) & = & \frac1{2\pi} \frac{v \mu_0}{2 D L}.
\edm
We are interested in late times and, thus, the terms involving $e^{\omega_- t}$ can be neglected. Therefore,
\be
\left< z \right>(t\to\infty) = i 2 \pi \left[ \frac d{dk_\parallel}b_+ + b_+ t \frac d{dk_\parallel}\omega_+ \right]_{k_\parallel=0}.
\ee
Moreover, we neglect terms which are higher order than linear in $1/L$ and, thus, we obtain
\bdm
& & \left< z \right>(t\to\infty)\nonumber\\
& = & i 2 \pi \left[ -i\frac{v\mu_0}{4\pi D}-i\frac{v^2}{24\pi D^2L}+i\frac{v^2}{12\pi DL} t \right]\nonumber\\
& = & \frac{v\mu_0}{2D} +\frac{v^2}{12 D^2L}- \frac{v^2}{6DL} t.
\label{ex_z_late}
\edm
We can average this result over $\mu_0$ and only consider the term involving a factor of $t$ to derive
\be
\int_{-1}^{+1} d\mu_0 \; \left< z \right> (t \to \infty) = - \frac{v^2}{6DL} t = - \frac{\kappa_\parallel}{L} t
\ee
which agrees with the result given by Eq.  (\ref{eq:mean}).

By using Eqs. (\ref{superpos}) and (\ref{eq:exzn}) for $n=2$, we obtain
\bdm
\left< z^2 \right> & = & - 2 \pi \Bigg[ e^{\omega_+t}\frac{d^2}{d k_\parallel^2}b_+ +2te^{\omega_+t}\frac{d}{d k_\parallel}\omega_+\frac d{dk_\parallel}b_+\nonumber\\
&+ & b_+te^{\omega_+t}\frac{d^2}{dk_\parallel^2}\omega_++b_+e^{\omega_+t}t^2\left(\frac{d}{dk_\parallel}\omega_+\right)^2\nonumber\\
& + & e^{\omega_-t}\frac{d^2}{d k_\parallel^2}b_- +2te^{\omega_-t}\frac{d}{d k_\parallel}\omega_-\frac d{dk_\parallel}b_-\nonumber\\
& + & b_-e^{\omega_-t}t\left(\frac{d}{dk_\parallel}\omega_-\right)^2+b_-t^2e^{\omega_-t}\frac{d^2}{dk_\parallel^2}\omega_-\Bigg]_{k_\parallel=0}.\nonumber\\
\edm
Again, we are interested in late times, and thus we obtain
\bdm
& & \left<z^2\right>(t\to\infty)\nonumber\\
& = & -2\pi\Bigg[ \Bigg(2\frac{d}{d k_\parallel}\omega_+\frac d{dk_\parallel}b_+ +b_+\frac{d^2}{dk_\parallel^2}\omega_+\Bigg)t \nonumber\\
& + & b_+t^2\left(\frac{d}{dk_\parallel}\omega_+\right)^2\Bigg]_{k_\parallel=0}.
\edm
Therein we use Eqs. (\ref{domegadk}) and (\ref{dbdk}) as well as
\be
\left[ \frac {d^2}{dk_\parallel^2} \omega_+ \right]_{k_\parallel=0} = \frac{v^2}{3} \frac{v^2 - 12 D^2 L^2}{12 D^3 L^2}.
\ee
Neglecting terms that are higher order than quadratic in $1/L$ gives us
\bdm
\left< z^ 2\right>(t\to\infty) & = & \left(\frac{v^2}{3D}-\frac{v^3\mu_0}{3 D^2L}-\frac{v^4}{18 D^3L^2}\right)t \nonumber\\
& + & \frac{v^4}{36 D^2L^2}t^2.
\label{ex_z2_late}
\edm
Combining Eqs.  (\ref{ex_z_late}) and (\ref{ex_z2_late}) yields
\be
\left[ \big< z^2 \big> - \big< z \big>^2 \right]_{t \to \infty} = \left[ \frac{v^2}{3D} - \frac{v^3\mu_0}{6D^2L} - \frac{v^4}{36D^3L^2}\right]t.\nonumber\\
\ee
Next, if one takes the $\mu_0-$average of this result, they derive
\be
\left[ \big< z^2 \big> - \big< z \big>^2 \right]_{t \to \infty} = 2 \frac{v^2}{6D} \left( 1 - \frac{v^2}{12 D^2 L^2} \right) t
\ee
the same expression as given by Eq. (\ref{zMSD}).
\section{The $N$-Dimensional Subspace Approximation}\label{hybrid}

In the following we discuss the $N$-dimensional subspace approximation. This approach is in detail discussed in Shalchi (2024) for the Fokker-Planck equation without
adiabatic focusing. In the current section we extend this method to include the focusing effect. Furthermore, we shall demonstrate that the $10$-dimensional subspace
approximation provides a very accurate approximation in most cases. This approach has to be understood as a semi-analytical method because some steps have to be
performed with a computer.

Within the $N$-dimensional subspace approximation, we employ the expansion in Legendre polynomials as given by Eq. (\ref{expandF}). The time-dependent functions
$C_n (t)$ therein are obtained via solving numerically the matrix equation provided by Eq. (\ref{cn_exact}). The initial conditions $C_n (0)$ in the latter equation are obtained
via Eq. (\ref{InitialCoeffRelation}). The matrix $\boldsymbol{M}$ in our matrix equation is provided by Eq. (\ref{MatrixwithElements}) where all matrix elements are given
as the coefficient of $C_n$ in Eq. (\ref{CRelations}). In principle $\boldsymbol{M}$ is an infinite matrix. However, as an approximate, we replace this matrix by an
$N \times N$ matrix by cutting off the expansion in Eq. (\ref{expandF}). We call this an $N$-dimensional subspace approximation. Pure analytical treatments for $N=2$
and $N=3$ are presented in the current paper but numerically we can consider any value for $N$. Motivated by Shalchi (2024) we consider the case $N=10$ and refer
to it as the $10$-dimensional subspace approximation.

To test the validity of various subspace approximations, we also solve the focused transport equation numerically. This is done by employing an \textit{implicit Euler method}.
The numerical methodology employed here aligns with the one used in Shalchi (2024). It is a straightforward process to incorporate the focusing term. First, we employ the
variable transformations
\be
\tilde{t} = D t \quad\textnormal{and}\quad \tilde{z} = \frac{Dz}{v}
\ee
which transform Eq.  (\ref{FPeq}) into
\be
\frac{\d f}{\d \tilde{t}} +\mu \frac{\d f}{\d \tilde{z}} = \frac\d{\d\mu} \left[ \left( 1 - \mu^2 \right) \frac{\d f}{\d\mu} \right] - \frac12 \xi \left( 1 - \mu^2 \right) \frac{\d f}{\d\mu}
\label{FourierFPeqTrans}
\ee
with the dimensionless parameter
\be
\xi = \frac{v}{D L}.
\ee

In the following subsections, we shall compare the subspace approximation method for $2$, $3$, and $10$-dimensions with the numerical solution. Pure analytical
results for the case $N=3$ are presented in Appendix \ref{sec:3dim}. We provide results for $\xi = 0.2$, $1$, and $5$ corresponding to different strengths of the
focusing effect. Moreover, at the end of this section, we compare the computation times needed to obtain the solution with the aforementioned methods.

\begin{figure}[H]
\centering
\includegraphics[scale=0.5]{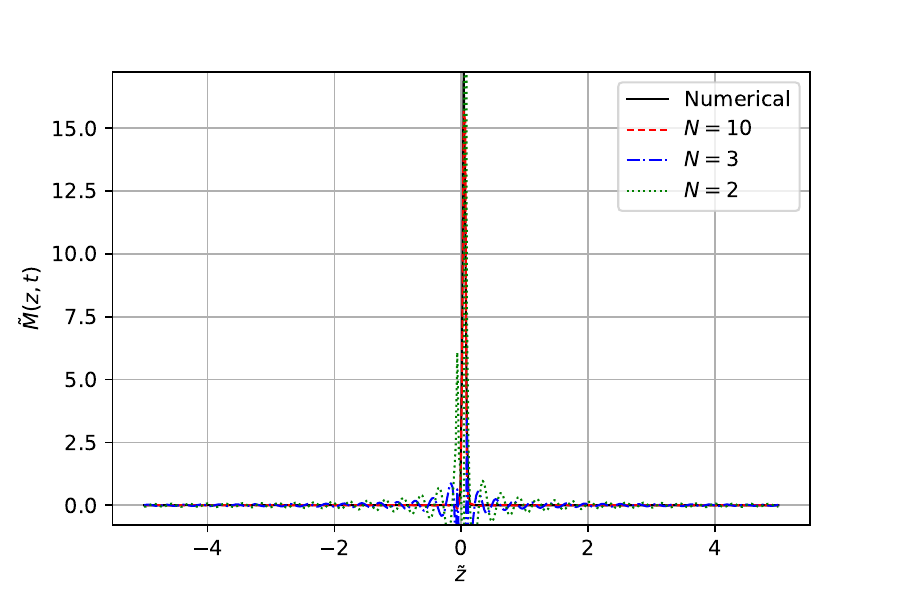}
\caption{The graph compares the $\mu$-integrated solution of the subspace method for various dimensions and the pure numerical solution at the time $\tilde{t} = D t = 0.1$
for an initial value of $\mu_0 = 0.5$. For the focusing parameter we set $\xi = 0.2$.}
\label{mu_integrated_5_01}
\end{figure}

\begin{figure}[H]
\centering
\includegraphics[scale=0.5]{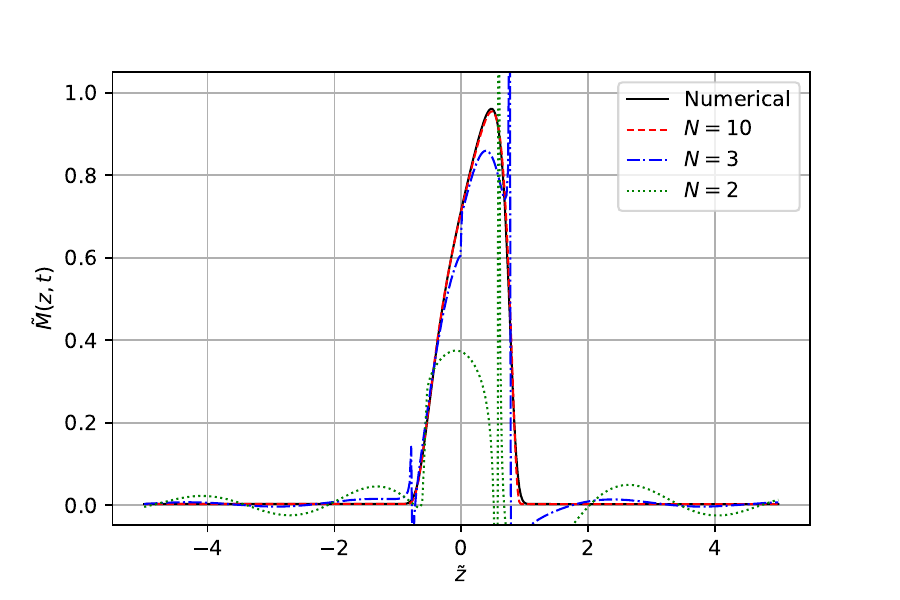}
\caption{Caption is as in Figure \ref{mu_integrated_5_01} except that we have used $\tilde{t} = 1$.}
\label{mu_integrated_5_1}
\end{figure}

\begin{figure}[H]
\centering
\includegraphics[scale=0.5]{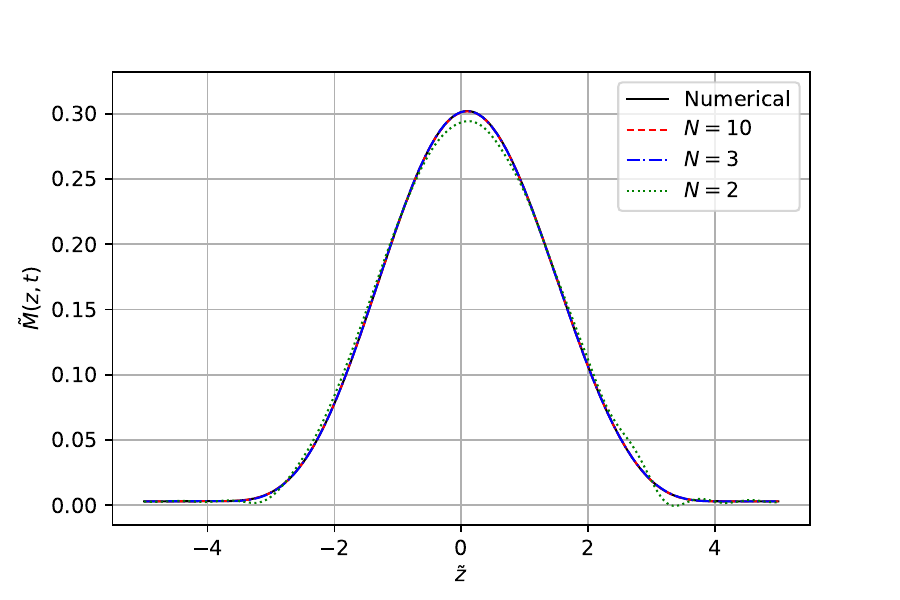}
\caption{Caption is as in Figure \ref{mu_integrated_5_01} except that we have used $\tilde{t} = 5$.}
\label{mu_integrated_5_5}
\end{figure}

First, we consider the case $\xi = 0.2$. Figures \ref{mu_integrated_5_01}-\ref{mu_integrated_5_5} illustrate the $\mu$-integrated solution at the different times
$\tilde{t} = D t = 0.1$, $1$, and $5$, respectively. Moreover, we use an initial value of $\mu_0 = 0.5$ as an example. Recall the expression for the $\mu$-integrated
distribution function $M(t,z)$ is given by Eq. (\ref{sub_mu_integral}). For early times we can see the used sharp initial conditions. At intermediate times the particles
move away from the their initial position. Due to the finite particle propagation speed, the distribution function is zero for $\left| \tilde{z} \right| > \tilde{t}$. For late
times the distribution function becomes Gaussian.

\begin{figure}[H]
\centering
\includegraphics[scale=0.5]{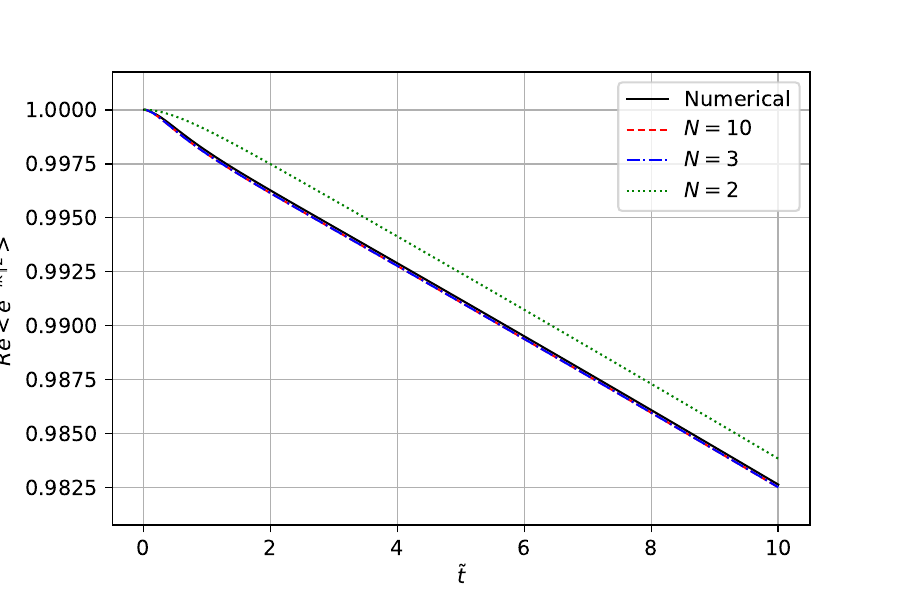}
\caption{Comparison of the $\mu$-averaged characteristic function between the subspace method for various dimensions and the numerical solution. We have used an initial
value of $\mu_0 = 0$ and for the focusing parameter we have set $\xi = 0.2$. Here we have kept the dimensionless wave number $\tilde{k} = v k / D = 0.1$ constant and varied time.}
\label{char_5_k_01}
\end{figure}

\begin{figure}[H]
\centering
\includegraphics[scale=0.5]{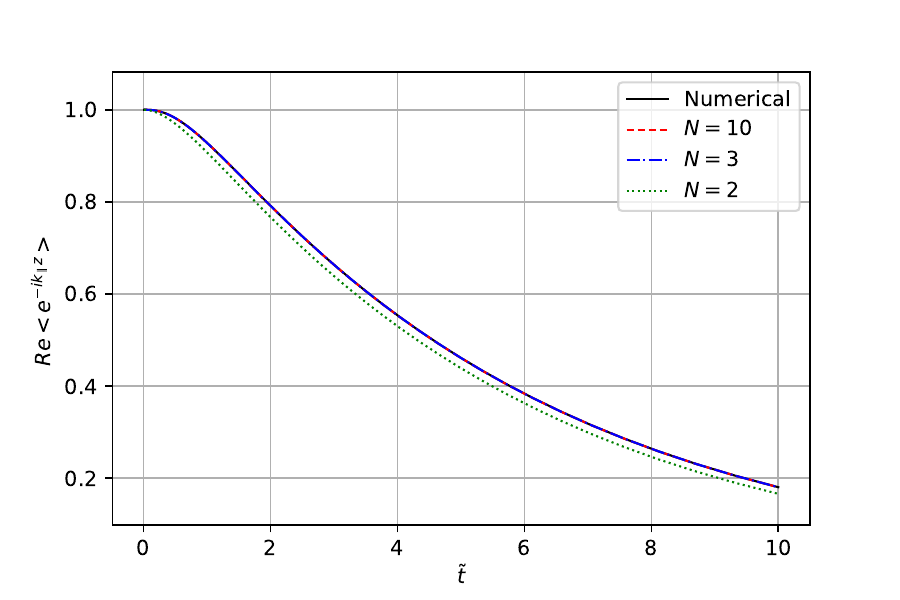}
\caption{Caption is as in Figure \ref{char_5_k_01} except that we have used $\tilde{k} = 1$.}
\label{char_5_k_1}
\end{figure}

\begin{figure}[H]
\centering
\includegraphics[scale=0.5]{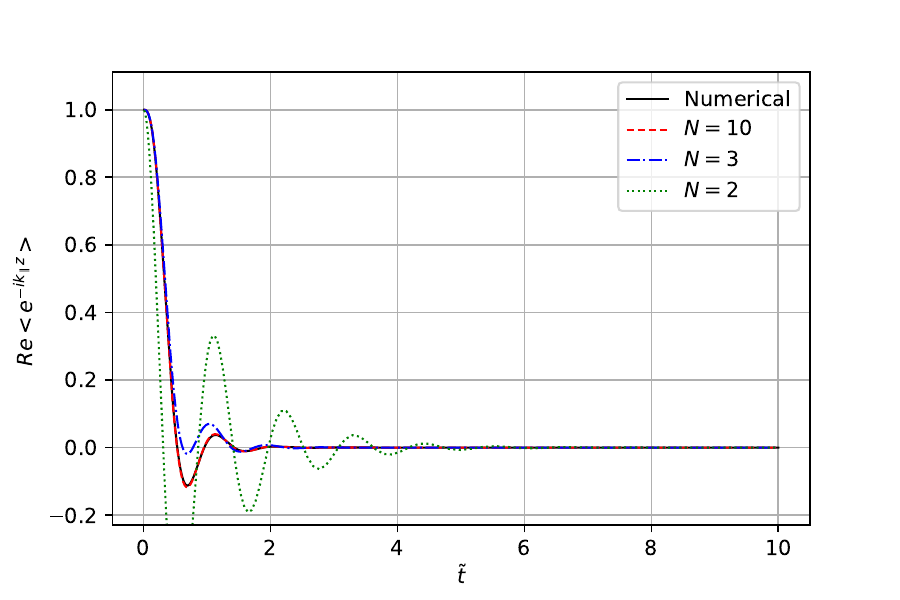}
\caption{Caption is as in Figure \ref{char_5_k_01} except that we have used $\tilde k = 10$.}
\label{char_5_k_10}
\end{figure}

Next, we examine the $\mu$-averaged characteristic function obtained via
\be
\left< e^{- i k_{\parallel} z} \right> = \frac{1}{2} \int_{-1}^{+1} d \mu 
\int_{-\infty}^{+\infty} d z \; f \big( z, \mu, t \big) e^{- i k_{\parallel} z}.
\ee
It follows from Section \ref{char_function} that
\be
\left< e^{- i k_{\parallel} z} \right> =2\pi C_0(k_\parallel,t).
\ee
Figures \ref{char_5_k_01}-\ref{char_5_k_10} demonstrate the accuracy of the subspace method for the real component of the characteristic function. We keep the wave number
constant and vary time.

\begin{figure}[H]
\centering
\includegraphics[scale=0.5]{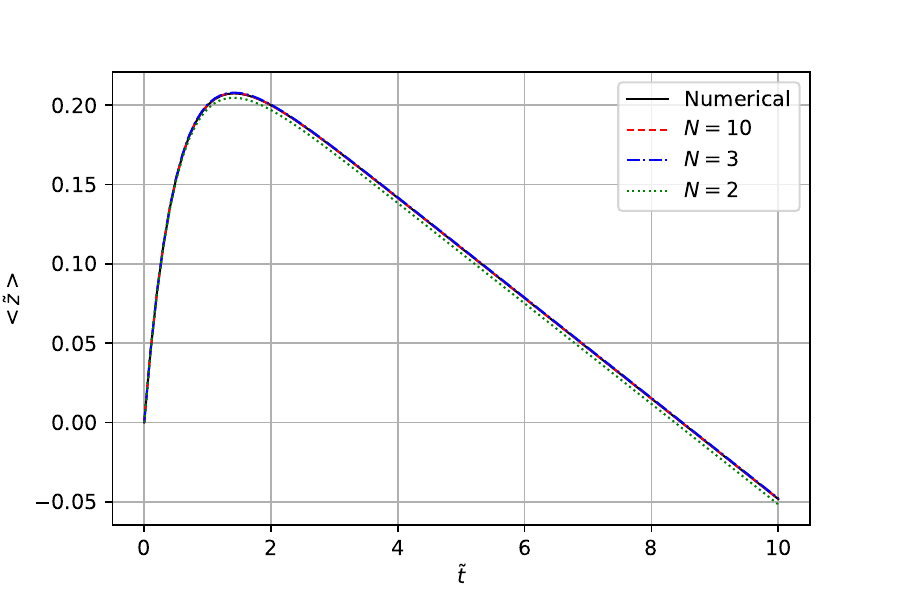}
\caption{Comparison of the mean position $\left< \tilde{z} \right>$ as a function of time $\tilde{t}$ for the $N$-dimensional subspace method and the numerical solution.
Here, we have used $\mu_0 = 0.5$ and $\xi = 0.2$.}
\label{Expectzxi15}
\end{figure}

\begin{figure}[H]
\centering
\includegraphics[scale=0.5]{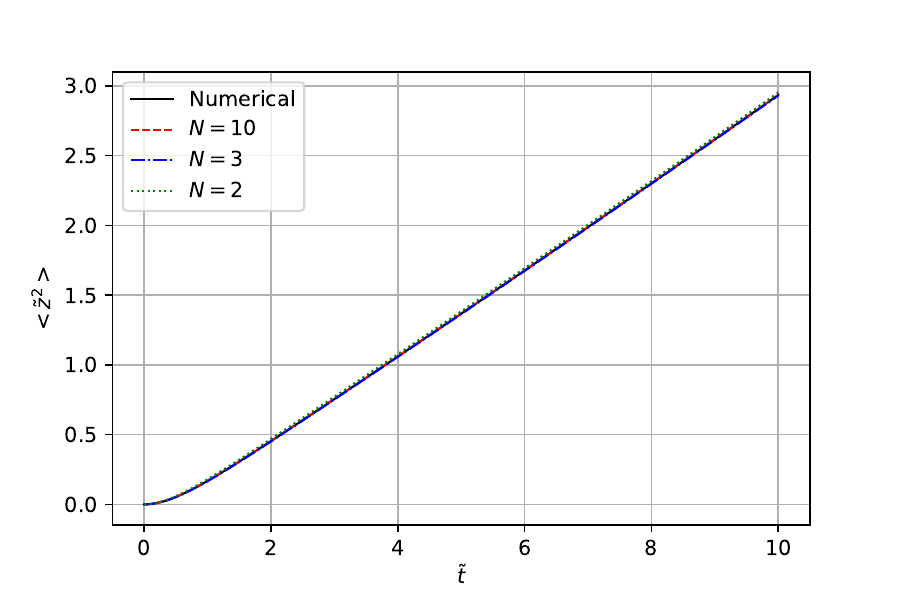}
\caption{Comparison of the second moment $\left< \tilde{z}^2 \right>$ as a function of time $\tilde{t}$ for the $N$-dimensional subspace method and the numerical solution.
Here, we have used $\mu_0 = 0.5$ and $\xi = 0.2$. }
\label{MSDxi15}
\end{figure}

In Figure \ref{Expectzxi15} we compare results for the expectation value $\left< \tilde{z} \right>$ obtained by employing Eq.  (\ref{eq:exzn}) for $n=1$. In
Figure \ref{MSDxi15} we compare the mean square displacements $\left< \tilde{z}^2 \right>$. The latter quantity is obtained by using Eq. (\ref{eq:exzn}) for $n=2$.

\begin{figure}[H]
\centering
\includegraphics[scale=0.5]{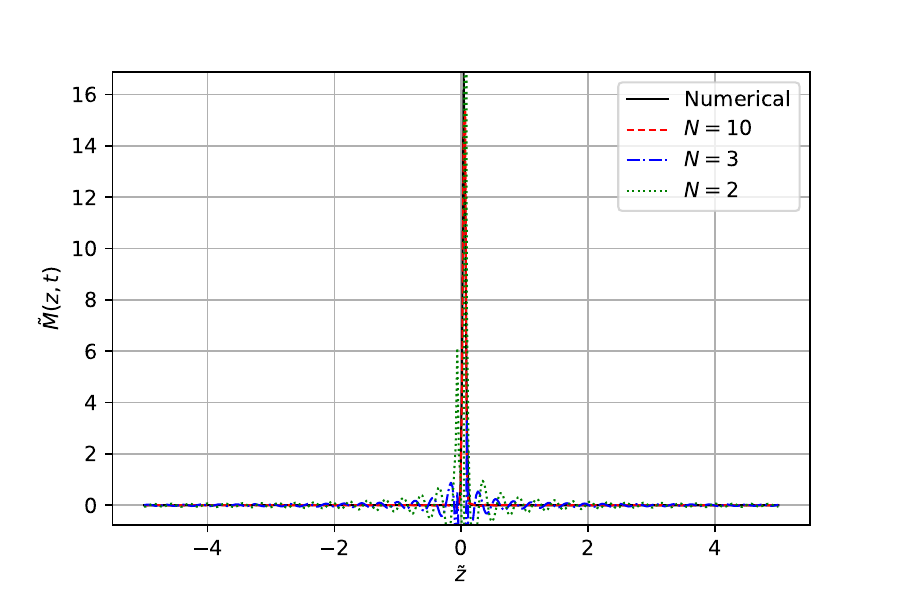}
\caption{This graph compares the $\mu$-integrated solution of the subspace method for various dimensions and the numerical solution at the time $\tilde{t} = 0.1$
for an initial value of $\mu_0 = 0.5$. Here we have set $\xi = 1$.}
\label{mu_integrated_1_01}
\end{figure}

\begin{figure}[H]
\centering
\includegraphics[scale=0.5]{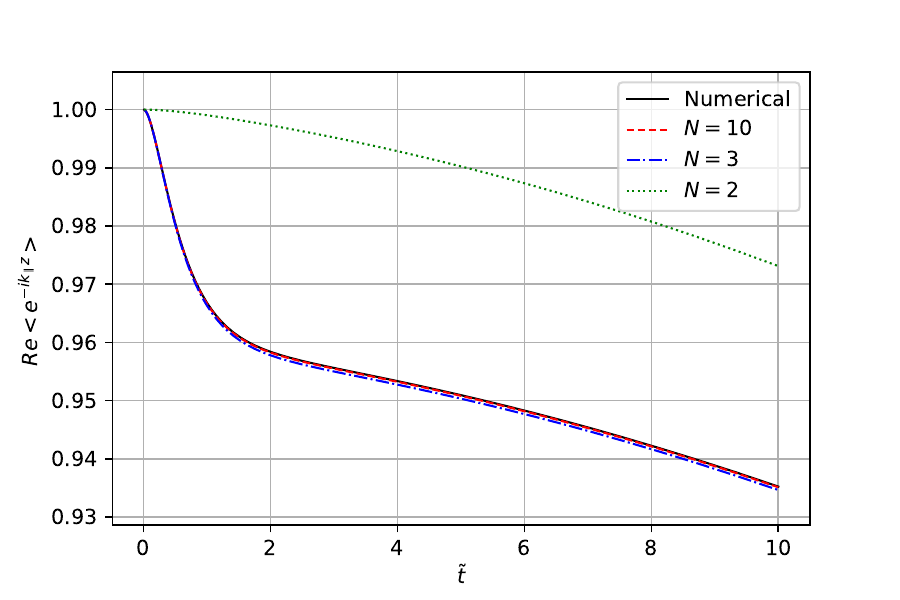}
\caption{Comparison of the $\mu$-averaged characteristic function between the subspace method for various dimensions and the numerical solution. We have used an initial
value of $\mu_0 = 0$ and set $\xi =1$. We kept the dimensionless wave number $\tilde{k} = v k / D = 0.1$ constant and varied time.}
\label{char_1_k_01}
\end{figure}

\begin{figure}[H]
\centering
\includegraphics[scale=0.5]{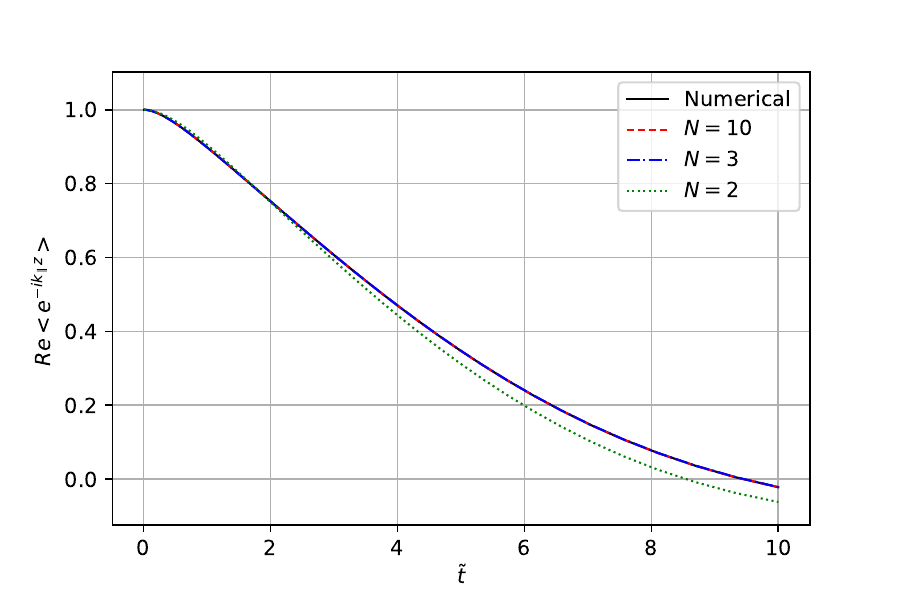}
\caption{Caption is as in Figure \ref{char_1_k_01} but we have set $\tilde{k} = 1$.}
\label{char_1_k_1}
\end{figure}

\begin{figure}[H]
\centering
\includegraphics[scale=0.5]{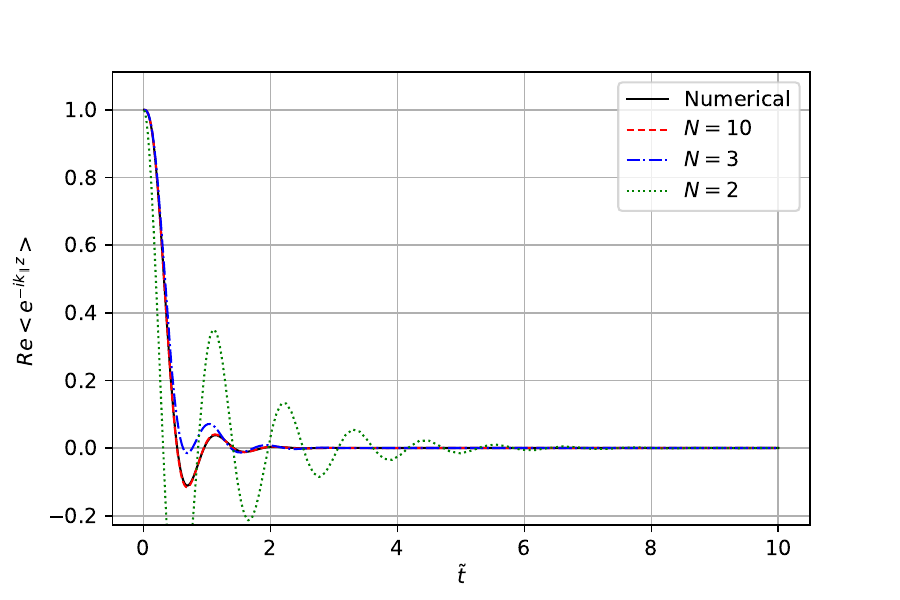}
\caption{Caption is as in Figure \ref{char_1_k_01} but we have set $\tilde{k} = 10$.}
\label{char_1_k_10}
\end{figure}

\begin{figure}[H]
\centering
\includegraphics[scale=0.5]{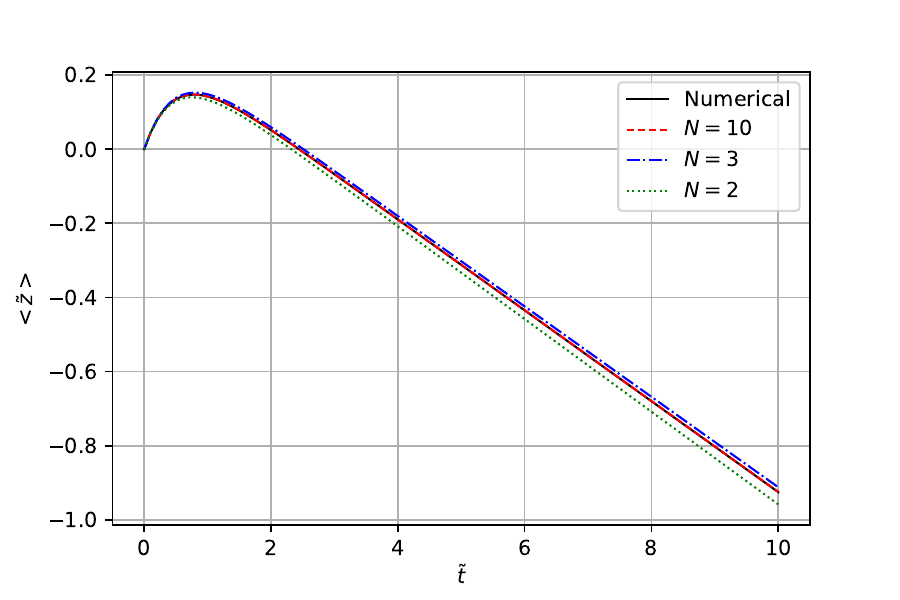}
\caption{Comparison of the mean position $\left< \tilde{z} \right>$ as a function of time $\tilde{t}$ for the $N$-dimensional subspace method and the numerical solution.
Here, we have set $\mu_0=0.5$ and $\xi =1$.}
\label{Expectzxi1}
\end{figure}

\begin{figure}[H]
\centering
\includegraphics[scale=0.5]{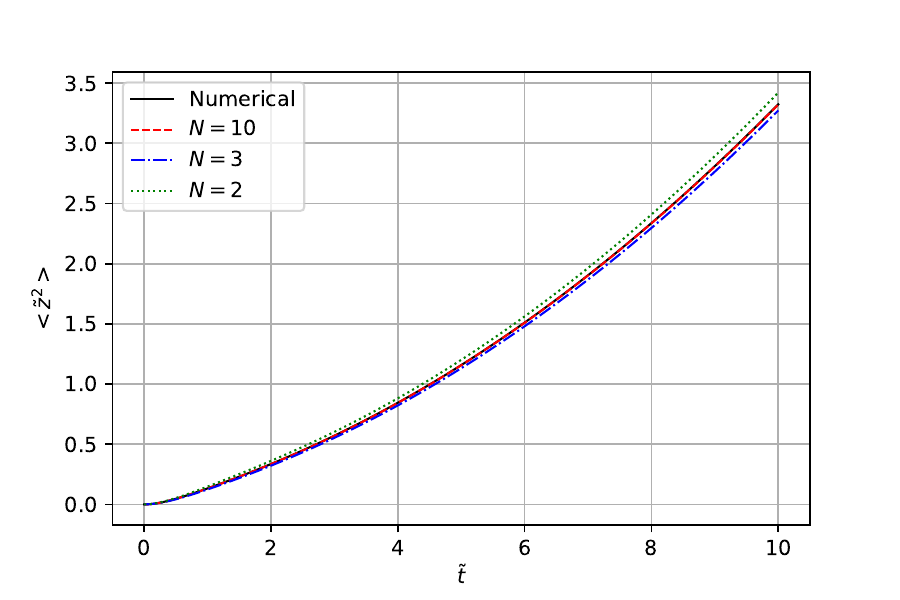}
\caption{Comparison of the second moment $\left< \tilde{z}^2 \right>$ as a function of time $\tilde{t}$ for the $N$-dimensional subspace method and the numerical solution.
Here, we have used $\mu_0 = 0.5$ and $\xi =1$. }
\label{MSDxi1}
\end{figure}

To explore the influence of the parameter $\xi$, we repeat our calculations by choosing the value $\xi = 1$. The different solutions are visualized via
Figures \ref{mu_integrated_1_01}-\ref{char_1_k_10} and the first and second moments are shown via Figures \ref{Expectzxi1} and \ref{MSDxi1}, respectively.

\begin{figure}[H]
\centering
\includegraphics[scale=0.5]{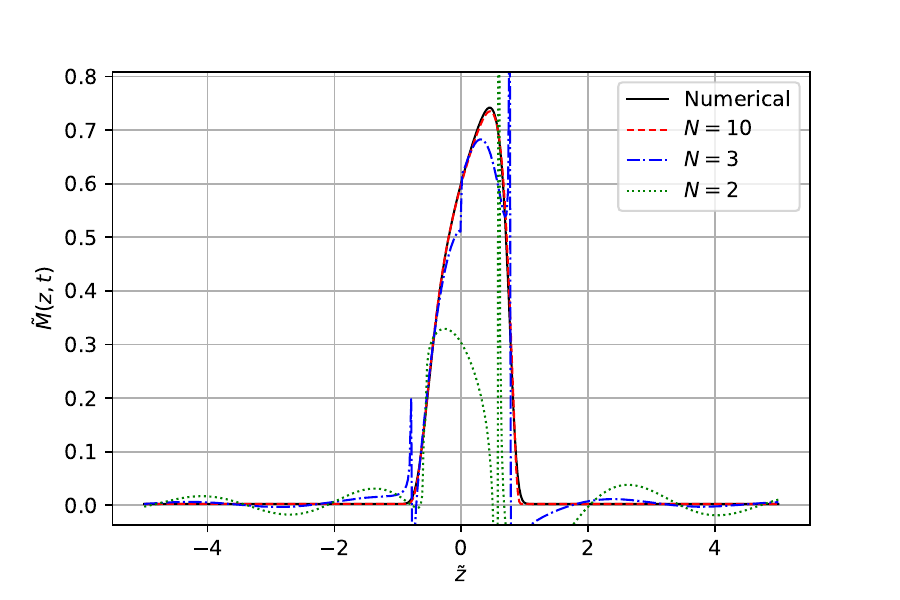}
\caption{Caption is as in Figure \ref{mu_integrated_1_01} except that we have used $\tilde{t} = 1$.}
\end{figure}

\begin{figure}[H]
\centering
\includegraphics[scale=0.5]{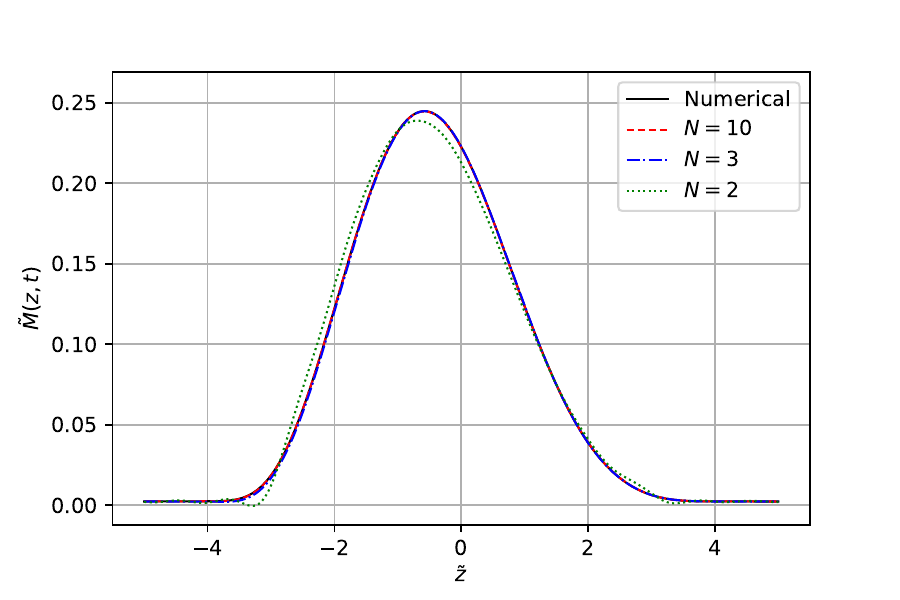}
\caption{Caption is as in Figure \ref{mu_integrated_1_01} except that we have used $\tilde{t} = 5$.}
\end{figure}

\begin{figure}[H]
\centering
\includegraphics[scale=0.5]{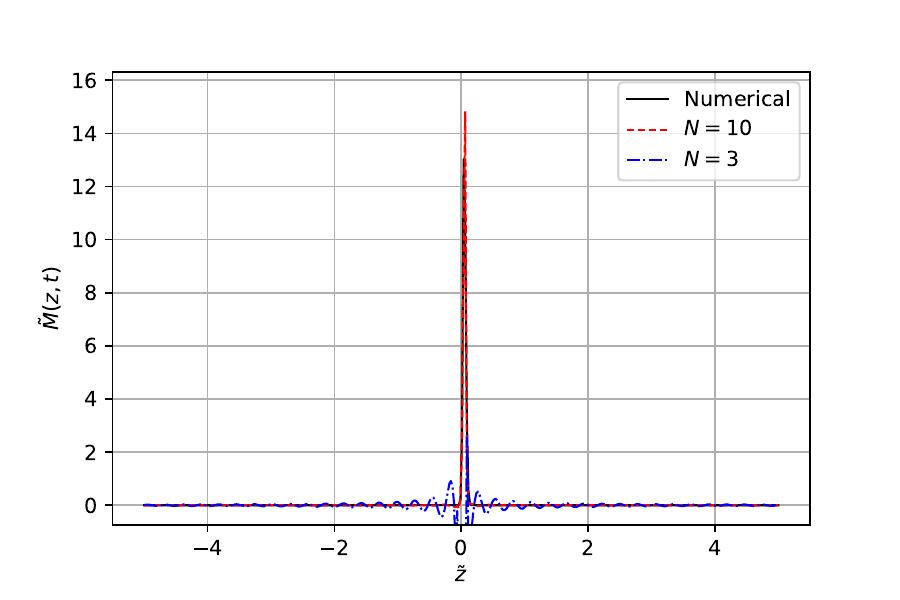}
\caption{This graph compares the $\mu$-integrated solution of the subspace method for various dimensions and the numerical solution at the time $\tilde{t} = 0.1$ for an initial
value of $\mu_0 = 0.5$. Here we have set $\xi = 5$.}
\label{mu_integrated_02_01}
\end{figure}

\begin{figure}[H]
\centering
\includegraphics[scale=0.5]{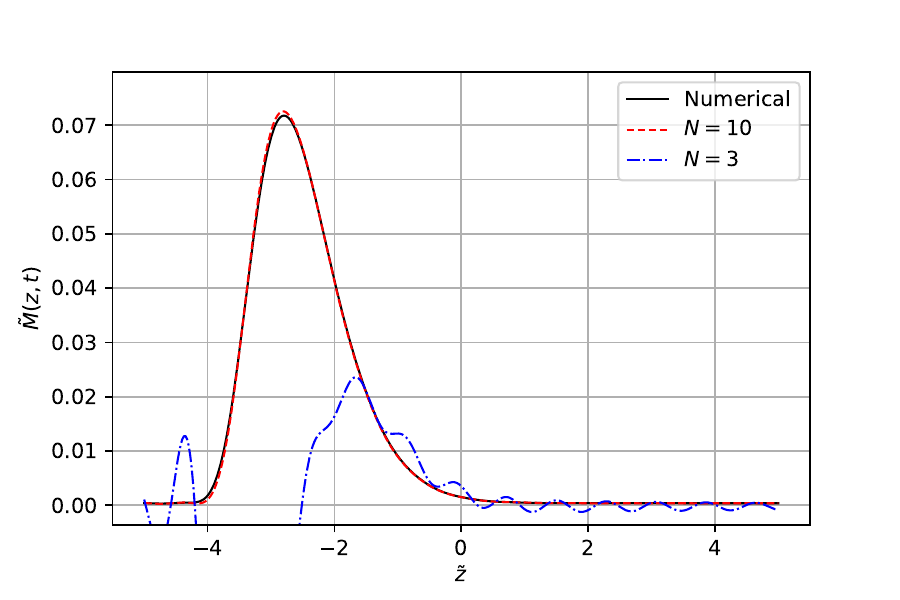}
\caption{Caption is as in Figure \ref{mu_integrated_02_01} but we have used $\tilde{t} = 5$.}
\label{mu_integrated_02_5}
\end{figure}

\begin{figure}[H]
\centering
\includegraphics[scale=0.5]{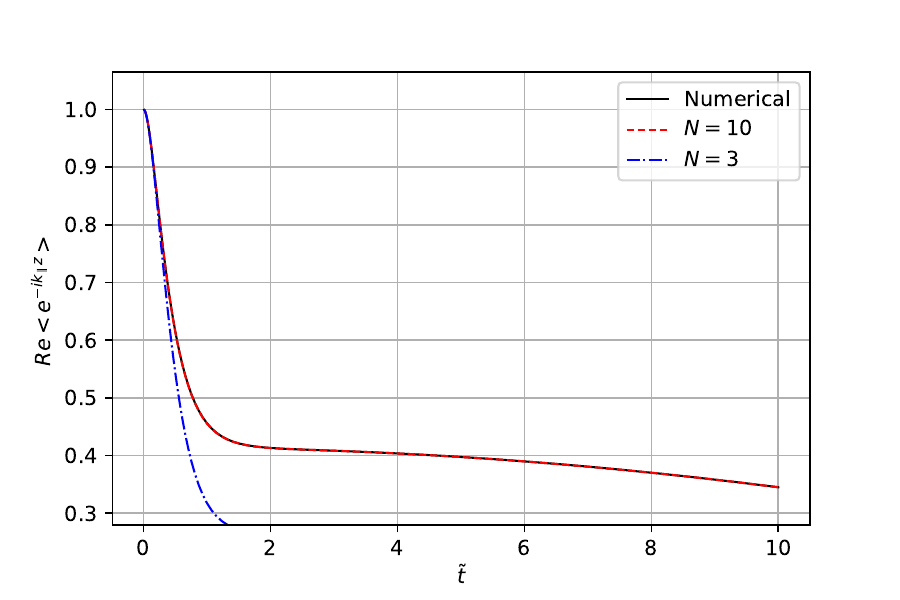}
\caption{Comparison of the $\mu$-averaged characteristic function between the subspace method for various dimensions and the numerical solution. Here we have used
an initial pitch-angle cosine of $\mu_0 = 0$ and for the focusing parameter we set $\xi = 5$. We have kept the dimensionless wave number $\tilde{k} = v k / D = 0.1$ constant
and varied time.}
\label{char_02_k_01}
\end{figure}

\begin{figure}[H]
\centering
\includegraphics[scale=0.5]{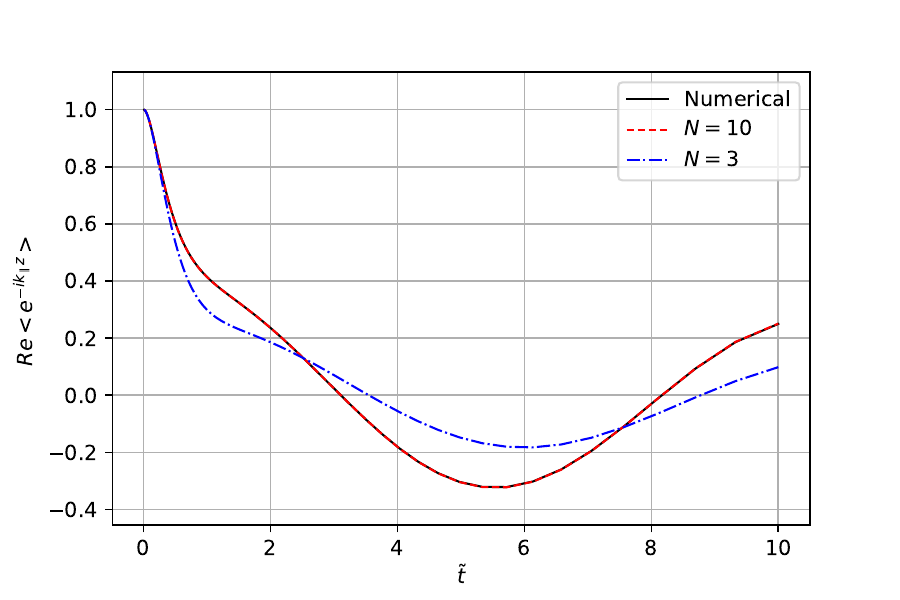}
\caption{Caption is as in Figure \ref{char_02_k_01} but we have set $\tilde{k} = 1$.}
\label{char_02_k_1}
\end{figure}

\begin{figure}[H]
\centering
\includegraphics[scale=0.5]{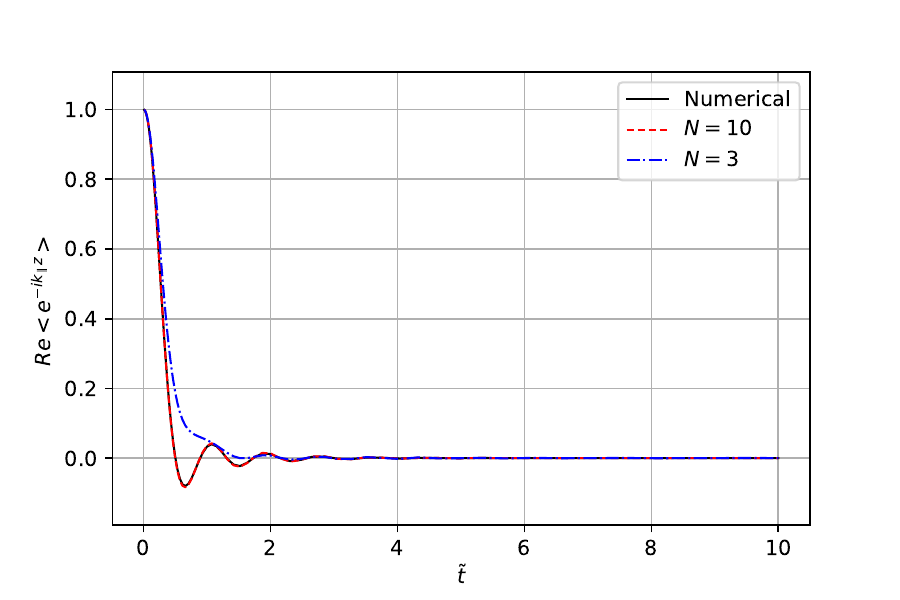}
\caption{Caption is as in Figure \ref{char_02_k_01} but we have set $\tilde{k} = 10$.}
\label{char_02_k_10}
\end{figure}

\begin{figure}[H]
\centering
\includegraphics[scale=0.5]{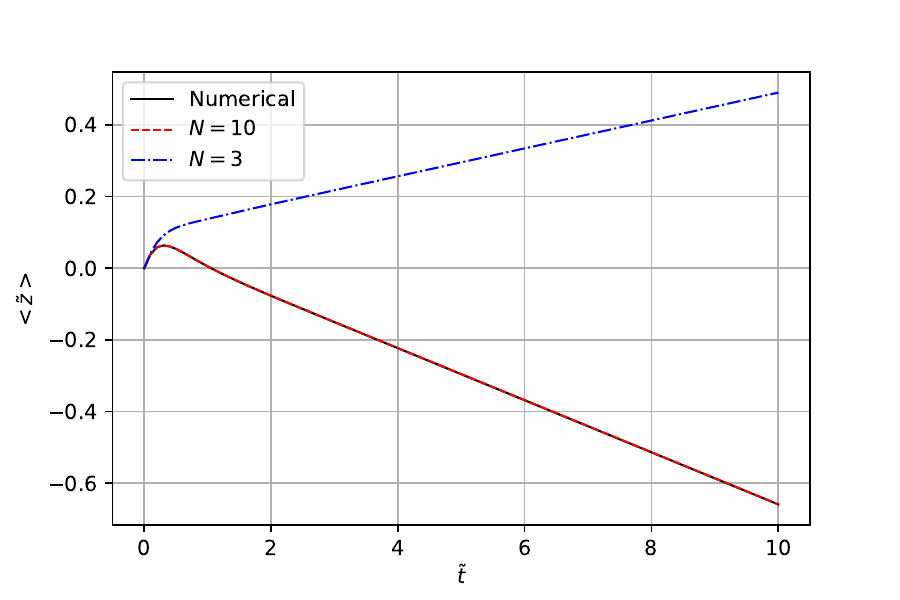}
\caption{Comparison of the mean position $\left< \tilde{z} \right>$ as a function of time $\tilde{t}$ for the $N$-dimensional subspace method and the numerical solution.
Here, we have set $\mu_0 = 0.5$ and $\xi = 5$.}
\label{Expectzxi5}
\end{figure}

\begin{figure}[H]
\centering
\includegraphics[scale=0.5]{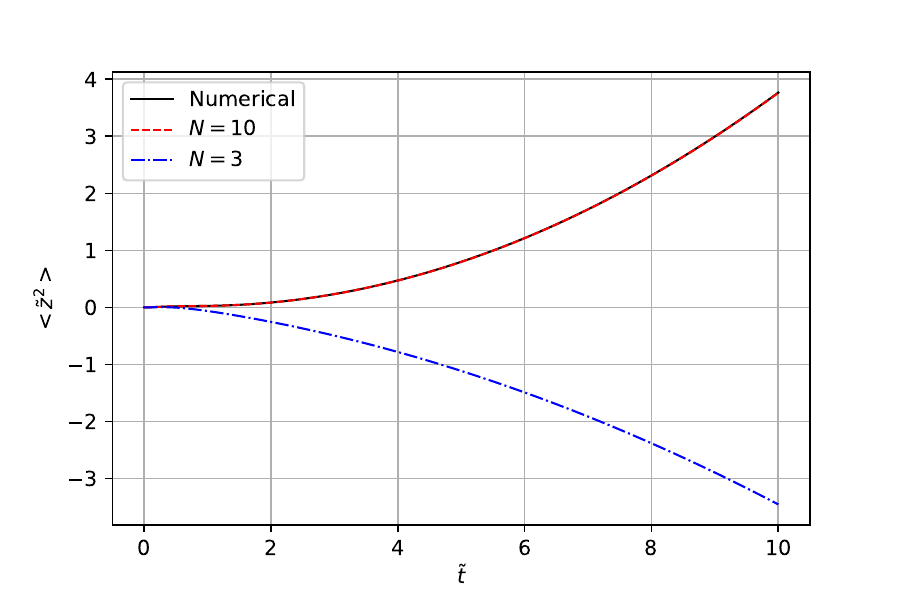}
\caption{Comparison of the moment $\left< \tilde{z}^2 \right>$ with respect to time $\tilde{t}$ for the $N$-dimensional subspace method and the numerical solution. Here, we have
used $\mu_0 = 0.5$ and $\xi = 5$. }
\label{MSDxi5}
\end{figure}

As last example we consider the case $\xi=5$. As we can see by considering Eq.  (\ref{pos_eig}), the solution does not converge for the two-dimensional subspace approximation.
Thus, we do not include this case in the following plots. We only compare the numerical solution with the cases $N=3$ and $N=10$. Figures \ref{mu_integrated_02_01}-\ref{char_02_k_10}
show the corresponding solutions and Figures \ref{Expectzxi5} and \ref{MSDxi5} depict the first and second moments.

\begin{figure}[H]
\centering
\includegraphics[scale=0.5]{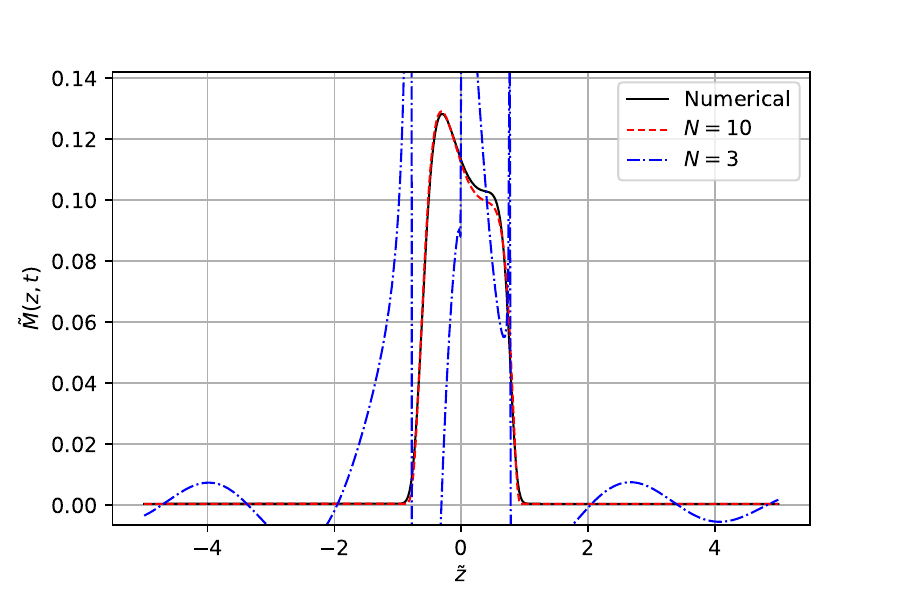}
\caption{Caption is as in Figure \ref{mu_integrated_02_01} but we have used $\tilde{t} = 1$.}
\end{figure}

\subsection{Time Comparisons}\label{time_comparing}

In the following we compare the computational time needed to obtain the solution using the $N$-dimensional subspace approximation with the time to find the solution numerically
by using implicit Euler integration. Tables \ref{time02}-\ref{time5} show the computational times for the different methods. We have only considered the time to compute each
$F_{k_\parallel}$; in other words, computing the integral in Eq.  (\ref{FourierTransform}) is not included. As the tables demonstrate, numerically computing the solution for early
times is marginally quicker, but for any other time, the subspace method is significantly faster. The cause of the numerical solution being quicker for early times is due to the fact
that in this case, very few time steps are required. Moreover, we also observe that the parameter $\xi$ does not affect the time required to compute the solution,  regardless of the
method used. It is important to note that each runtime was calculated using \textit{Python} on the same computer.

It should also be noted that for the subspace methods, we compute the solution for 100 values of $\mu$, whereas for the numerical solution we must use $2000$ values
to ensure the solution is accurate.

\begin{table*}
\centering
\begin{tabular}
{ |p{3cm}|p{3cm}|p{3cm}|p{3cm}|p{3cm}| }
 \hline
$Dt$ &2D &3D & 10D & Pure Numerical\\
\hline
 $0.1$ &$1112.7$ & $1739.3$&$6790.3$&$6116.4$\\
\hline
$1$&$34.7$&$54.9$&$213.4$&$1811.3$\\
\hline
$5$ &$3.0$&$4.8$&$18.4$&$778.8$\\
\hline
\end{tabular}
\caption{For different values of $Dt$, we compare the time in seconds to compute the solution using the subspace approximation for various values of $N$ with the numerical solution
based on implicit Euler integration. Here, we have have used $\xi = 0.2$.}
\label{time02}
\end{table*}

\begin{table*}
\centering
\begin{tabular}
{ |p{3cm}|p{3cm}|p{3cm}|p{3cm}|p{3cm}| }
 \hline
$Dt$ &2D &3D & 10D & Pure Numerical\\
\hline
 $0.1$&$1125.3$&$1757.1$&$6841.1$&$6113.9$\\
\hline
$1$&$35.3$&$55.7$&$217.3$&$1809.1$\\
\hline
$5$ &$3.0$&$4.8$&$19.0$&$775.8$\\
\hline
\end{tabular}
\caption{Caption is as in Table \ref{time02}, except that we have used $\xi = 1$.}
\label{time1}
\end{table*}

\begin{table*}
\centering
\begin{tabular}
{ |p{3cm}|p{3cm}|p{3cm}|p{3cm}| }
 \hline
$Dt$ &3D & 10D & Pure Numerical\\
\hline
$0.1$&$1767.6$&$6865.0$&$6102.3$\\
\hline
$1$&$57.2$&$220.5$&$1804.2$\\
\hline
$5$&$4.8$&$19.1$&$779.0$\\
\hline
\end{tabular}
\caption{Caption is as in Table \ref{time02}, except that we have used $\xi=5$.}
\label{time5}
\end{table*}

\subsection{Limitations of the $N$-Dimensional Subspace Approximation}\label{flaw}

\begin{figure}[H]
\centering
\includegraphics[scale=0.5]{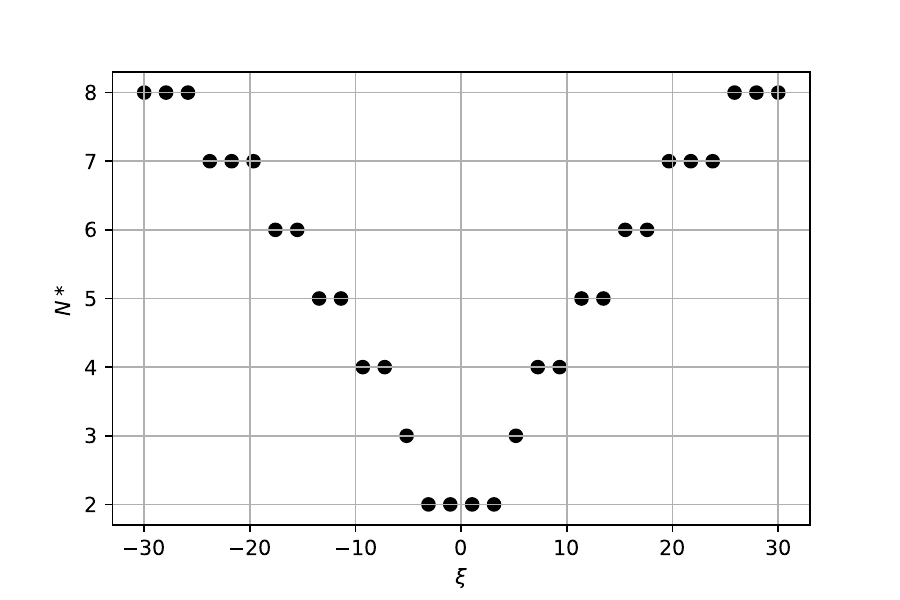}
\caption{The relation between $\xi$ and $\tilde N$, which is the minimum dimension one must use for the subspace method to produce a good approximation.}
\label{nstar_xi}
\end{figure}

As we discussed at the end of Sect. \ref{2dsection}, one fact that hinders the use of the $2$-dimensional subspace approximation is the possibility of the eigenvalues having a positive
real component. However, in this section, we show that this problem does not only occur when using two dimensions. In fact, for any given $N$-dimensional subspace approximation,
if the parameter $\xi = v/(DL)$ is sufficiently large, we encounter a similar issue. We illustrate this concept in Figure \ref{nstar_xi}. To understand this figure, we begin with observing
through numerical evidence that if $\{\omega_n(k)\}_{n=1}^N$ is the set of eigenvalues of the matrix in Eq. (\ref{eq:M}) associated with the $N$-dimensional subspace approximation,
we have that for each $1\leq n\leq N$,
\be
\lim_{k \rightarrow \infty} \Re \left\{ \omega_n(k) \right\} = \tilde\omega_n
\ee
where $\Re$ denotes the real component. In other words, the real component of each eigenvalue converges to some finite value as $k$ goes to infinity.

The flaw of the $N$-dimensional subspace approximation is that if we do not take $N$ to be large enough, there are values of $n$ such that $\tilde \omega_n$ is positive.
Figure \ref{nstar_xi} shows the relation between the parameter $\xi$, and the smallest value of $N$, denoted by $\tilde N$, which does not give rise to this issue. It should be noted
that we do not claim that for a given $\xi$, the $\tilde N$-dimensional subspace method guarantees a good approximation. Instead, we claim that if one wants a good approximation,
they must take $N$ to be at least $\tilde N$. For example, Figure \ref{nstar_xi} shows that for $\xi=5$ and $N=3$ one does not encounter this problem. However, as demonstrated
above, this does not give a good approximation.

There is another inconvenience that arises from the fact that the real component of the eigenvalues does not decay to zero. That is the uncertainty of knowing when to truncate the integral
in Eq. (\ref{FourierTransform}). For example, by solving Eq. (\ref{FourierFPeq}) numerically, it can be seen that the magnitude of $F_{k_\parallel}$ decays exponentially as the wave number
increases. This in turn leads to naturally truncating the integral in Eq. (\ref{FourierTransform}) when $F_{k_\parallel}$ is deemed to be sufficiently small. As for the subspace method however,
we do not achieve this decay, and thus deciding when to truncate the Fourier integral is not clear. Nevertheless, in our results for the $\mu$-integrated solution, we truncate the Fourier integral
for the subspace method at the same wave number as we did for the numerical solution. By judging the figures we produced above, this choice of truncation is seen to be sufficient.

\section{Summary and Conclusion}

The understanding of particle transport is a very fundamental problem in space physics and astrophysics. Furthermore, there are a variety of applications of transport theory results. Most important
examples are the theories of diffusive shock acceleration and solar modulation (see Zank et al. (2000a), Li et al. (2003), Zank et al. (2004), Li et al. (2005), Zank et al. (2006), Dosch \& Shalchi (2010),
Li et al. (2012), Ferrand et al. (2014), Hu et al. (2017), Shen \& Qin (2018), Engelbrecht \& Wolmarans (2020), Moloto \& Engelbrecht (2020), Engelbrecht \& Moloto (2021), Shen et al. (2021),
and Ngobeni et al. (2022)).

Using the $N$-dimensional subspace approximation method has been employed in previous literature to solve the Fokker-Planck equation (see, e.g., Zank et al. (2000b), Lasuik \& Shalchi (2019),
and Shalchi (2024)). In the current paper we extended the subspace method by including the effect of adiabatic focusing. We first used the $2$-dimensional subspace method to obtain approximate
results for the solution as well as several expectation values. Unexpectedly, we saw that some results within this method are independent of the focusing term, which shows that $2$-dimensions
are not sufficient here. Therefore, we developed a three-dimensional subspace approximation and showed that the corresponding results are more aligned with what is expected, as demonstrated
in Appendix \ref{sec:3dim}. On the other hand, we show that using $10$-dimensions provides an accurate, yet quick approximation. This semi-analytical approach provides an alternative
compared to pure numerical solutions which can be very time-consuming. However, Section \ref{flaw} discusses that if the focusing parameter $\xi = v / (DL)$ is sufficiently large, one must proceed
with caution and ensure that using $10$-dimensions is still sufficient.

A central question in the theory of focused transport is how the focusing effect influences the analytical form of the parallel spatial diffusion coefficient. There have been contradicting results but
it was demonstrated in Danos et al. (2013) that how the parallel diffusion coefficient looks like with focusing effect depends on how this diffusion coefficient is computed. Results which are obtained
by employing the TGK formula are different compared to the one found by using mean square displacments. This difference can also be seen in the current paper. Within the two-dimensional
subspace approximation we derived Eq. (\ref{defkappabar}) corresponding to a diffusion coefficient obtained via mean square displacement. Eq. (\ref{kappabarMSD3}) provides the same result
but based on the three-dimensional subspace approximation. It is slightly more accurate. However, if one uses the TGK formula for determining the parallel spatial diffusion coefficient with focusing
one obtains Eqs. (\ref{tgk}) and (\ref{TGKwith3D}), respectively. Since one is interested in the pitch-angle averaged distribution function and the corresponding diffusion-convection equation,
it seems to be more appropriate to use the diffusion coefficient given by Eqs. (\ref{defkappabar}) and (\ref{kappabarMSD3}). This also confirms the results originally obtained in Shalchi (2011b).

In the current paper we have used the standard approach to explore the parallel spatial diffusion coefficient with focusing effect. This means we determined this coefficient by using a given form
of the pitch-angle scattering coefficient $D_{\mu\mu}$. It should be noted, however, that there have been investigations of this scattering coefficient and how it is influenced by the focusing effect.
Tautz et al. (2014) as well as Florinski (2024) determined $D_{\mu\mu}$ with focusing effect by employing quasi-linear theory and they found a significant change of this scattering coefficient.
However, for such more realistic forms of $D_{\mu\mu}$ it is even more difficult to solve the focused transport equation.

In the current paper we have employed the subspace method to solve Eq. (\ref{FPeq}). The latter equation desribes a pitch-angle isotropization process like the same equation but without
focusing effect. However, Eq. (\ref{FPeq}) does not conserve the norm. An alternative focused transport equation can be derived by employing the transformation
\be
\tilde{f} \left( t, z, \mu \right) = f \left( t, z, \mu \right) e^{z/L}
\ee
leading to a \textit{modified focused transport equation} which does conserve the norm. In the sequel to the current paper we employ the subspace approximation to solve this modified
equation.

\begin{acknowledgments}
{\it Support by the Natural Sciences and Engineering Research Council (NSERC) of Canada is acknowledged.}
\end{acknowledgments}

\clearpage
\onecolumngrid
\appendix

\section{The Three-Dimensional Subspace Approximation}\label{sec:3dim}

In the following we derive analytical results using the three-dimensional subspace approximation. For the case $N=3$, Eq. (\ref{linear_system_cn}) becomes
\bdm
\left(
\begin{array}{c}
\dot{C}_0      \\[0.5cm]
\dot{C}_1      \\[0.5cm]
\dot{C}_2      \\
\end{array}
\right) = \left(
\begin{array}{ccc}
0   							\quad & - \frac{1}{3} i v k_{\parallel} - \frac{v}{3 L} 	\quad &	0																\\[0.5cm]
- i v k_{\parallel}		\quad & - 2 D														\quad & - \frac{2}{5} i v k_{\parallel}-\frac35\frac vL		\\[0.5cm]
0								\quad & - \frac{2}{3} i v k_{\parallel} + \frac{v}{3 L} \quad & - 6 D
\end{array}
\right) \left(
\begin{array}{c}
C_0      \\[0.5cm]
C_1 	 	\\[0.5cm]
C_2 	 	\\
\end{array}
\right).\label{eq:3dim}
\edm
The \textit{ansatz}
\be
C_n \big( t \big) = b_n e^{\omega t} \quad \textnormal{for} \quad n=0,1,2
\ee
leads to the matrix equation
\bdm
\omega \left(
\begin{array}{c}
b_0      \\[0.5cm]
b_1      \\[0.5cm]
b_2      \\
\end{array}
\right) = \left(
\begin{array}{ccc}
0   							\quad & - \frac{1}{3} i v k_{\parallel} - \frac{v}{3 L} 	\quad &	0																\\[0.5cm]
- i v k_{\parallel}		\quad & - 2 D														\quad & - \frac{2}{5} i v k_{\parallel}-\frac35\frac vL		\\[0.5cm]
0								\quad & - \frac{2}{3} i v k_{\parallel} + \frac{v}{3 L} \quad & - 6 D
\end{array}
\right) \left(
\begin{array}{c}
b_0    \\[0.5cm]
b_1 	 \\[0.5cm]
b_2 	 \\
\end{array}
\right)
\edm
corresponding to a simple eigenvalue problem. Alternatively, this can be written as
\bdm
\left(
\begin{array}{ccc}
-\omega 					\quad & - \frac{1}{3} i v k_{\parallel} - \frac{v}{3 L} 	\quad &	0																\\[0.5cm]
- i v k_{\parallel}		\quad & - 2 D	-\omega										\quad & - \frac{2}{5} i v k_{\parallel}-\frac35\frac vL		\\[0.5cm]
0								\quad & - \frac{2}{3} i v k_{\parallel} + \frac{v}{3 L} 	\quad & - 6 D-\omega
\end{array}
\right) \left(
\begin{array}{c}
b_0      \\[0.5cm]
b_1 	 \\[0.5cm]
b_2 	 \\
\end{array}
\right) = 0.\nonumber
\edm
Non-trivial solutions of the latter equation are obtained by setting the determinant of this $3 \times 3$ matrix equal to zero. We find
\be
\omega^3+8D\omega^2+\left[12D^2+\frac3{5}v^2k_\parallel^2+\frac{v^2}{5L^2}-i\frac35\frac{v^2k_\parallel}{L}\right]\omega+2D\left[v^2k_\parallel^2-i\frac{v^2k_\parallel}{L}\right]=0
\label{eq:cubic}
\ee
corresponding to a cubic equation for $\omega$.

\subsection{The Three-Dimensional Eigenvalues}

The solutions of Eq. (\ref{eq:cubic}) are
\be
\omega_{k}=-\frac13\left(8D+\sigma^kC+\frac{\triangle_0}{\sigma^kC}\right)\label{eq:3dimeig}
\ee
for $k=0 ,1 ,2$. Here, we have used
\be
\sigma = - \frac12 + i \frac{\sqrt3}2
\ee
and
\be
C=\left(\frac{\triangle_1\pm\sqrt{\triangle_1^2-4\triangle_0^3}}2\right)^{1/3}
\ee
where either sign can be chosen. Furthermore, we have used
\be
\triangle_0=28D^2-\frac35\left(3v^2k_\parallel^2+\frac{v^2}{L^2}-i3\frac{v^2k_\parallel}{L}\right),\label{eq:triangle0}
\ee
and
\be
\triangle_1=160D^3+\frac{18}5D\left[3v^2k_\parallel^2-4\frac{v^2}{L^2}-i3\frac{v^2k_\parallel}{L}\right].\label{eq:triangle1}
\ee

\subsection{The Three-Dimensional Initial Conditions}

We can use Eq. (\ref{InitialCoeffRelation}) to easily derive
\bdm
C_0 \big( t = 0 \big) & = & \frac{1}{2 \pi}, \nonumber\\
C_1 \big( t = 0 \big) & = & \frac{3 \mu_0}{2 \pi}, \nonumber\\
C_2 \big( t = 0 \big) & = & \frac{5}{4 \pi} \big( 3 \mu_0^2 - 1 \big).
\label{eq:3diminitial}
\edm
For convenience we write Eq. (\ref{eq:3dim}) as
\bdm
\left(
\begin{array}{c}
\dot{C}_0      \\[0.5cm]
\dot{C}_1      \\[0.5cm]
\dot{C}_2      \\
\end{array}
\right) = \left(
\begin{array}{ccc}
0   			\quad\quad & M_{01}	\quad\quad &	0				\\[0.5cm]
M_{10}		\quad\quad & M_{11}	\quad\quad & M_{12}		\\[0.5cm]
0				\quad\quad & M_{21}	\quad\quad & M_{22}
\end{array}
\right) \left(
\begin{array}{c}
C_0   \\[0.5cm]
C_1 	 \\[0.5cm]
C_2 	 \\
\end{array}
\right).
\label{eq:3dimsimp}
\edm
Now we assume
\be
C_0=b_0e^{\omega_0t}+b_1e^{\omega_1t}+b_2e^{\omega_2t}
\label{eq:3dimc0}
\ee
where $b_0$, $b_1$, and $b_2$ are unknown constants. From Eq. (\ref{eq:3dimsimp}), we obtain
\be
C_1=\frac{1}{M_{01}}\dot C_0 =\frac1{M_{01}}\left[\omega_0b_0e^{\omega_0t}+\omega_1b_1e^{\omega_1t}+\omega_2b_2e^{\omega_2t}\right].
\label{eq:3dimc1}
\ee
Eq. (\ref{eq:3dimsimp}) also provides
\bdm
C_2 & = & \frac1{M_{12}}\left[\dot C_1-M_{10}C_0-M_{11}C_1\right]\nonumber\\
& = &\frac1{M_{12}}\Bigg[\left(\frac{\omega_0^2}{M_{01}}-M_{10}-\frac{M_{11}\omega_0}{M_{01}}\right)b_0e^{\omega_0t}+\left(\frac{\omega_1^2}{M_{01}}-M_{10}-\frac{M_{11}\omega_1}{M_{01}}\right)b_1e^{\omega_1t}\nonumber\\
& + & \left(\frac{\omega_2^2}{M_{01}}-M_{10}-\frac{M_{11}\omega_2}{M_{01}}\right)b_2e^{\omega_2t}\Bigg].\label{eq:3dimc2}
\edm
Combining Eqs. (\ref{eq:3dimc0})-(\ref{eq:3dimc2}) with Eq.  (\ref{eq:3diminitial}) leads to the three equations
\bdm
 \frac{1}{2 \pi}&=&b_0+b_1+b_2, \nonumber\\
\frac{3 \mu_0}{2 \pi}M_{01} & = & \omega_0b_0+\omega_1b_1+\omega_2b_2, \nonumber\\
\frac{5}{4 \pi} \big( 3 \mu_0^2 - 1 \big)M_{12} & = &\left(\frac{\omega_0^2}{M_{01}}-M_{10}-\frac{M_{11}\omega_0}{M_{01}}\right)b_0+\left(\frac{\omega_1^2}{M_{01}}-M_{10}-\frac{M_{11}\omega_1}{M_{01}}\right)b_1\nonumber\\
& + & \left(\frac{\omega_2^2}{M_{01}}-M_{10}-\frac{M_{11}\omega_2}{M_{01}}\right)b_2.
\label{eq:3dimb}
\edm
Using the first two equations, the third line of Eq. (\ref{eq:3dimb}) can be rewritten as
\be
\frac{5}{4 \pi} \big( 3 \mu_0^2 - 1 \big) M_{12} M_{01} + \frac1{2\pi} M_{10} M_{01} + \frac{3\mu_0}{2\pi} M_{11} M_{01} = \omega_0^2 b_0 + \omega_1^2 b_1 + \omega_2^2 b_2.
\label{eq:3dimbsimp}
\ee
The obtained system of equations has the solutions
\bdm
b_0&=&\frac{\omega_1\omega_2+\left(\frac vL+ivk_\parallel\right)(\omega_1+\omega_2)\mu_0+\frac{5}{2 } M_{12}M_{01}\big( 3 \mu_0^2 - 1 \big)+M_{10}M_{01}+3M_{11}M_{01}\mu_0}{2\pi(\omega_0-\omega_1)(\omega_0-\omega_2)},\nonumber\\
b_1&=&\frac{\omega_0\omega_2 +\left(\frac vL+ivk_\parallel\right)(\omega_0+\omega_2)\mu_0+\frac{5}{2 } M_{12}M_{01}\big( 3 \mu_0^2 - 1 \big)+M_{10}M_{01}+3M_{11}M_{01}\mu_0}{2\pi(\omega_1-\omega_0)(\omega_1-\omega_2)},\nonumber\\
b_2&=&\frac{\omega_0\omega_1+\left(\frac vL+ivk_\parallel\right)(\omega_0+\omega_1)\mu_0+\frac{5}{2 }M_{12}M_{01} \big( 3 \mu_0^2 - 1 \big)+M_{10}M_{01}+3M_{11}M_{01}\mu_0}{2\pi(\omega_2-\omega_0)(\omega_2-\omega_1)},
\edm
corresponding to formulas for the three needed coefficients.

\subsection{The Limit $k_\parallel = 0$}

In the case of $k_\parallel=0$, the eigenvalues are easy to find. Here,  Eq. (\ref{eq:cubic}) simplifies to
\be
\omega\left[\omega^2+8D\omega+12D^2+\frac15\frac{v^2}{L^2}\right]=0
\ee
and thus
\bdm
\omega_0 (k_\parallel=0) & = & 0,\nonumber\\
\omega_1 (k_\parallel=0) & = & -4D - \sqrt{4D^2-\frac15\frac{v^2}{L^2}},\nonumber\\
\omega_2 (k_\parallel=0) & = & -4D + \sqrt{4D^2-\frac15\frac{v^2}{L^2}}.
\label{eq:omega0}
\edm
Moreover,  we have
\bdm
b_0 (k_\parallel=0) & = & \frac{\omega_1(k_\parallel=0)\omega_2(k_\parallel=0)+\frac vL(\omega_1(k_\parallel=0)+\omega_2(k_\parallel=0))\mu_0+\frac{1}{2 } \frac{v^2}{L^2}\big( 3 \mu_0^2 - 1 \big)+2\frac{Dv}L\mu_0}{2\pi\omega_1(k_\parallel=0)\omega_2(k_\parallel=0)},\nonumber\\
b_1 (k_\parallel=0) & = & \frac{\frac vL\omega_2(k_\parallel=0)\mu_0+\frac{1}{2 } \frac{v^2}{L^2}\big( 3 \mu_0^2 - 1 \big)+2\frac{Dv}L\mu_0}{2\pi\omega_1(k_\parallel=0)(\omega_1(k_\parallel=0)-\omega_2(k_\parallel=0))},\nonumber\\
b_2 (k_\parallel=0) & = & \frac{\frac vL\omega_1(k_\parallel=0)\mu_0+\frac{1}{2 }\frac{v^2}{L^2}\big( 3 \mu_0^2 - 1 \big)+2\frac{Dv}L\mu_0}{2\pi\omega_2(k_\parallel=0)(\omega_2(k_\parallel=0)-\omega_1(k_\parallel=0))}.\label{eq:b0}
\edm
As we shall see later, an important quantity is $C_1(k_\parallel=0,t)$. By Eqs. (\ref{eq:3dimc1}), (\ref{eq:omega0}), and (\ref{eq:b0}), we have
\bdm
C_1(k_\parallel,t=0) & = & -3\Bigg[\frac{\omega_2(k_\parallel=0)\mu_0+2D\mu_0+\frac{1}{2 } \frac{v}{L}\big( 3 \mu_0^2 - 1 \big)}{2\pi \left[ \omega_1(k_\parallel=0)-\omega_2(k_\parallel=0) \right]} e^{\omega_1(k_\parallel=0)t}\nonumber\\
& + & \frac{\omega_1(k_\parallel=0)\mu_0+2D\mu_0+\frac{1}{2 }\frac{v}{L}\big( 3 \mu_0^2 - 1 \big)}{2\pi \left[ \omega_2(k_\parallel=0)-\omega_1(k_\parallel=0) \right]}e^{\omega_2(k_\parallel=0)t}\Bigg].
\label{eq:c10}
\edm
This formula will be used in Section \ref{TheExpectValuemu} to compute the expectation value $\left< \mu \right>$.

\subsection{Expectation Values}

In the main part of this paper we employed the two-dimensional subspace approximation to compute several expectation values. In the following we determine the same quantities
but employ the three-dimensional subspace approximation. This is particularly important because several expectation values are independent of the focusing effect when the
two-dimensional approach is employed. We shall see that one advantage of using the extra dimension is that each result contains the focusing length $L$.

\subsubsection{The Characteristic Function}

The first few steps for calculating the characteristic function are provided by Eqs. (\ref{eq:characteristic})-(\ref{eq:Fint}) implying that
\bdm
\left< e^{- i k_{\parallel} z} \right> & = & \pi \int_{-1}^1 d\mu_0 \; C_0 (t)\nonumber\\
& = & \pi \int_{-1}^1d\mu_0 \left[b_0e^{\omega_0t}+b_1e^{\omega_1t}+b_2e^{\omega_2t}\right]\nonumber\\
& = &\frac{\omega_1\omega_2-\frac13v^2k_\parallel^2+i\frac13\frac{v^2k_\parallel}{L}}{(\omega_0-\omega_1)(\omega_0-\omega_2)}e^{\omega_0t}+\frac{\omega_0\omega_2-\frac13v^2k_\parallel^2+i\frac13\frac{v^2k_\parallel}{L}}{(\omega_1-\omega_0)(\omega_1-\omega_2)}e^{\omega_1t}+\frac{\omega_0\omega_1-\frac13v^2k_\parallel^2+i\frac13\frac{v^2k_\parallel}{L}}{(\omega_2-\omega_0)(\omega_2-\omega_1)}e^{\omega_2t}.
\label{eq:char_3dim}
\edm
In what follows, we simplify this result for small wave numbers and late times. To do this, we write $\omega=a+bk_\parallel+ck_\parallel^2$ with the constants $a,b,c$ independent of $k_\parallel$,
and place that into the eigenvalue equation (\ref{eq:cubic}) to obtain
\bdm
a^3+8Da^2&&+12D^2a+\frac{v^2}{5L^2}a+\left(3a^2b+16Dab+12D^2b+\frac15\frac{v^2}{L^2}b-i\frac35\frac{v^2}La-i2\frac{Dv^2}L \right)k_\parallel\nonumber\\
&&\qquad+\left(3ab^2+3a^2c+8Db^2+16Dac+12D^2c+\frac35v^2a+2Dv^2+\frac15\frac{v^2}{L^2}c-i\frac35\frac{v^2}Lb \right)k_\parallel^2=0\label{eq:eig_smallk}
\edm
where we neglect terms higher order than quadratic in $k_\parallel$. The equation found here has to be satisfied for any value of $k_\parallel$. Therefore, the terms without $k_\parallel$ must
be zero and we find
\be
a^3 + 8 D a^2 + 12 D^2 a + \frac{v^2}{5L^2} a = 0.
\ee
From this condition we can derive the following three values for the constant $a$:
\be
a=0 \quad\quad\textnormal{and}\quad\quad a= - 4 D \pm \sqrt{4 D^2 - \frac{v^2}{5 L^2}}.
\ee
Since we are interested in the late time limit, we can see that the term with the slowest decay in Eq. (\ref{eq:char_3dim}) corresponds to the eigenvalue with $a=0$. Henceforth, we set $a=0$
in Eq. (\ref{eq:eig_smallk}) to simplify it to
\be
\left(12D^2b+\frac15\frac{v^2}{L^2}b-i2\frac{Dv^2}L \right)k_\parallel+\left(8Db^2+12D^2c+2Dv^2+\frac15\frac{v^2}{L^2}c-i\frac35\frac{v^2}Lb \right)k_\parallel^2=0.
\label{eq:eig_smallk_2}
\ee
Now,  from the linear term,  we easily obtain the expression
\be
b = 10 i \frac{\frac{v^2}{DL}}{60+\frac{v^2}{D^2L^2}}.
\label{smallk_b}
\ee
If we place this into the quadratic term of Eq.  (\ref{eq:eig_smallk_2}),  we obtain
\be
c=-\frac{40\left(900\frac{v^2}D-25\frac{v^4}{D^3L^2}+\frac{v^6}{D^5L^4}\right)}{\left(60+\frac{v^2}{D^2L^2}\right)^3}.\label{smallk_c}
\ee
By combining Eqs.  (\ref{smallk_b}) and (\ref{smallk_c}),  we derive an expression for one eigenvalue,  which we denote by $\omega_0$:
\be
\omega_0=-\frac{40\left(900\frac{v^2}D-25\frac{v^4}{D^3L^2}+\frac{v^6}{D^5L^4}\right)}{\left(60+\frac{v^2}{D^2L^2}\right)^3}k_\parallel^2+i10\frac{\frac{v^2}{DL}}{60+\frac{v^2}{D^2L^2}}k_\parallel.
\ee
Since the factor $e^{\omega_0t}$ has significantly slower decay than the other exponential factors in Eq.  (\ref{eq:char_3dim}),  we can simplify it to
\be
\left< e^{- i k_{\parallel} z} \right>\approx\frac{\omega_1\omega_2-\frac13v^2k_\parallel^2+i\frac13\frac{v^2k_\parallel}{L}}{(\omega_0-\omega_1)(\omega_0-\omega_2)}e^{\omega_0t}.
\ee
Next,  we use the fact that $\omega_0\to 0$ as $k_\parallel\to0$ once more to simplify the factor
\be
\frac{\omega_1\omega_2-\frac13v^2k_\parallel^2+i\frac13\frac{v^2k_\parallel}{L}}{(\omega_0-\omega_1)(\omega_0-\omega_2)} \approx1.
\ee
If we assume further that $\xi = v/(DL)$ is small, we can additionally simplify Eqs. (\ref{smallk_b}) and (\ref{smallk_c}) to
\be
\qquad b= i\frac16 \frac{v^2}{DL} \quad\textnormal{and}\quad c = -\frac16\frac{v^2}D+\frac{7}{540}\frac{v^4}{D^3L^2},
\ee
resulting in
\be
\left< e^{- i k_{\parallel} z} \right>\approx e^{- \kappa_\parallel k_\parallel^2 t + \frac{7}{15}\frac{\kappa_\parallel^2 k_\parallel^2 t}{DL^2} + i\frac{\kappa_\parallel}{L}k_\parallel t}.
\label{charfunction3D}
\ee
Here, we have used the parallel diffusion coefficient without focusing as given by Eq. (\ref{eq:kparallel}). From Eq. (\ref{charfunction3D}) we can read off the parallel diffusion coefficient
with focusing effect and we find
\be
\bar{\kappa}_{\parallel} = \kappa_\parallel \left( 1 - \frac{7}{15} \frac{\kappa_\parallel}{D L^2} \right).
\ee
Therein we can use the diffusion coefficient without focusing and write
\be
\bar{\kappa}_{\parallel} = \kappa_\parallel \left( 1 - \frac{7}{90} \frac{v^2}{D^2 L^2} \right).
\ee
Or, we use the parallel mean free path without focusing (see Eq. (\ref{MFPnofocusing}) of the current paper) to write this as
\be
\bar{\kappa}_{\parallel} = \kappa_\parallel \left( 1 - \frac{14}{45} \frac{\lambda_{\parallel}^2}{L^2} \right).
\label{kappabarMSD3}
\ee
This result can be compared with Eq. (\ref{defkappabar}) which is based on the two-dimensional subspace approximation. Due to $14/45 \approx 1/3$ the results for the parallel
diffusion coefficient do not change significantly when going from the two-dimensional method to the three-dimensional one.

\subsubsection{The Velocity Correlation Function}

By following the ideas used in Eqs. (\ref{eq:velocitycorrelation})-(\ref{eq:velocitycorrsimp}), we have for the velocity correlation function
\be
V_{zz}(t) = \pi \frac{v^2}3 \int_{-1}^1 d \mu_0 \; \mu_0 C_1(k_\parallel=0,t).
\ee
With Eq. (\ref{eq:c10}), this becomes
\bdm
V_{zz}(t) & = & -\frac{v^2}3\left[\frac{\omega_2(k_\parallel=0)+2D}{\omega_1(k_\parallel=0)-\omega_2(k_\parallel=0)}e^{\omega_1t} + \frac{\omega_1(k_\parallel=0)+2D}{\omega_2(k_\parallel=0)-\omega_1(k_\parallel=0)}e^{\omega_2t}  \right]\nonumber\\
& = & \frac{ v^2}{6}\Bigg[\left(1-\frac{1}{\sqrt{1-\frac1{20}\frac{v^2}{D^2L^2}}}\right) e^{\omega_1(k_\parallel=0)t}+\left(1+\frac{1}{\sqrt{1-\frac1{20}\frac{v^2}{D^2L^2}}}\right) e^{\omega_2(k_\parallel=0)t} \Bigg].
\label{eq:vzzmain}
\edm
If we assume that the parameter $\xi$ is small, we can simplify this result to
\be
V_{zz} (t)=\frac{v^2}6\left[-\frac1{40}\frac{v^2}{D^2L^2} e^{\left(-6D+\frac1{20}\frac{v^2}{DL^2}\right)t}+\left(2+\frac1{20}\frac{v^2}{D^2L^2}\right)e^{\left(-2D-\frac1{20}\frac{v^2}{DL^2}\right)t}\right].
\ee
Integrating this over all times yields
\bdm
\int_{0}^\infty dt \; V_{zz}(t) & = & \frac{ v^2}{6}\Bigg[-\left(1-\frac{1}{\sqrt{1-\frac1{20}\frac{v^2}{D^2L^2}}}\right) \frac1{\omega_1(k_\parallel=0)}-\left(1+\frac{1}{\sqrt{1-\frac1{20}\frac{v^2}{D^2L^2}}}\right) \frac1{\omega_2(k_\parallel=0)} \Bigg]\nonumber\\
& = & \frac{ v^2}{6}\Bigg[-\left(\frac1{\omega_1(k_\parallel=0)}+\frac1{\omega_2(k_\parallel=0)}\right) +\frac{1}{\sqrt{1-\frac1{20}\frac{v^2}{D^2L^2}}}\left(\frac1{\omega_1(k_\parallel=0)}- \frac1{\omega_2(k_\parallel=0)}\right) \Bigg]\nonumber\\
& = & \frac{v^2}{6} \frac{1}{\omega_1 (k_\parallel=0) \omega_2 (k_\parallel=0)} \Bigg[-\left(\omega_1(k_\parallel=0)+\omega_2(k_\parallel=0)\right)\nonumber\\
& + & \frac{1}{\sqrt{1-\frac1{20}\frac{v^2}{D^2L^2}}}\left(\omega_2(k_\parallel=0)- \omega_1(k_\parallel=0)\right) \Bigg]\nonumber\\
& = & \frac{ v^2}{6}\frac{1}{12D^2+\frac15\frac{v^2}{L^2}}\Bigg[8D +\frac{1}{\sqrt{1-\frac1{20}\frac{v^2}{D^2L^2}}}4D\sqrt{1-\frac1{20}\frac{v^2}{D^2L^2}} \Bigg]\nonumber\\
& = & \frac{ v^2}{6}\frac{12D}{12D^2+\frac15\frac{v^2}{L^2}}.
\edm
In the weak focusing limit the corresponding parallel diffusion coefficient becomes
\be
\bar{\kappa}_{\parallel}^{TGK} = \int_{0}^\infty dt \; V_{zz} (t) = \frac{v^2}{6 D} \left( 1 - \frac{v^2}{60 D^2 L^2} \right).
\ee
Note, with the parallel diffusion coefficient without focusing effect (see again Eq. (\ref{eq:kparallel})) this can be written as
\be
\bar{\kappa}_{\parallel}^{TGK} =  \kappa_\parallel \left( 1 - \frac{1}{15} \frac{\lambda_{\parallel}^2}{L^2} \right).
\label{TGKwith3D}
\ee
This result can be compared with the time-integrated velocity correlation function based on the two-dimensional subspace approximation (see Eq. (\ref{tgk}) of the current paper).
While the two-dimensional approximation did not provide a result depending on the focusing effect, the formula based on the three-dimensional approximation predicts a
reduction of the parallel diffusion coefficient. Still, we can see a difference between diffusion coefficients based on mean square displacements (see Eq. (\ref{kappabarMSD3}))
and the one obtained be employing the TGK formula. This difference can always be observed and is a special feature of particle transport with focusing (see Danos et al. (2013)).

\subsubsection{The Expectation Value $\left< \mu \right>$}\label{TheExpectValuemu}

To determine the expectation value $\left< \mu \right>$, we employ Eq. (\ref{eq:exmu}). With Eq.  (\ref{eq:c10}), this becomes
\bdm
\left<\mu\right>&=& -\Bigg[\frac{\omega_2(k_\parallel=0)\mu_0+2D\mu_0+\frac{1}{2 } \frac{v}{L}\big( 3 \mu_0^2 - 1 \big)}{\omega_1(k_\parallel=0)-\omega_2(k_\parallel=0)} e^{\omega_1(k_\parallel=0)t}\nonumber\\
&\qquad&\qquad\qquad\qquad\qquad + \frac{\omega_1(k_\parallel=0)\mu_0+2D\mu_0+\frac{1}{2 }\frac{v}{L}\big( 3 \mu_0^2 - 1 \big)}{\omega_2(k_\parallel=0)-\omega_1(k_\parallel=0)}e^{\omega_2(k_\parallel=0)t}\Bigg].
\edm
Again, we assume that $\xi$ is small, and thus we obtain the simplified expression
\bdm
\left< \mu \right> & = & -\Bigg[\left(-\frac18\frac v{DL}(3\mu_0^2-1)+\frac{v^2}{80D^2L^2}\mu_0\right)e^{(-6D+\frac{v^2}{20DL^2})t}\nonumber\\
&\qquad&\qquad\qquad\qquad+\left(\frac18 \frac v{DL}(3\mu_0^2-1)-\mu_0-\frac{v^2}{80D^2L^2}\mu_0\right)e^{(-2D-\frac{v^2}{20DL^2})t}\Bigg].
\edm
Keeping in mind that the focusing effect is assumed to be weak here, we can easily see that
\be
\left< \mu \right>(t=0) =\mu_0.
\ee
Moreover,  we also have
\be
\left< \mu \right>_{t \rightarrow \infty} \rightarrow 0
\ee
regardless of what the value of the initial pitch-angle cosine $\mu_0$ is. This is what we have expected due to the pitch-angle isotropization process described by Eq. (\ref{FPeq}).

{}

\end{document}